\documentclass{JHEP3}
\usepackage{epsfig}  
\usepackage{amsmath}
\usepackage{amsfonts}
\usepackage{amssymb}
\usepackage{rotate}
\usepackage{xy}
\xyoption{matrix}\xyoption{arrow}\xyoption{frame}

\newcommand{\rfe}[1]{(\ref{eq:#1})}
\newcommand{\p}{\partial}
\newcommand{\bp}{{\bar\partial}}
\newcommand{\tr}{{\rm tr}~ }
\newcommand{\comment}[1]{}

\newcommand{\mapr}{\mathop{\longrightarrow}\limits}
\newcommand{\ul}{\underline}

\def\bracomplex#1{\left[\! #1 \!\right]}
\def\complex#1{#1}

\newcommand{\IIa}{{\rm IIa}}
\newcommand{\IIb}{{\rm IIb}}
\newcommand{\grade}{\varphi}
\newcommand{\dual}{*} 
\newcommand{\im}{{\rm im}}
\newcommand{\Hom}{{\rm Hom}}
\newcommand{\Ext}{{\rm Ext}}
\newcommand{\Mod}{{\rm Mod}}
\newcommand{\Coh}{{\rm Coh}}
\newcommand{\Quiv}{{\rm Quiv}}
\newcommand{\Cat}{{\rm Cat}}
\newcommand{\rank}{{\rm rank~}}

\newcommand{\CF}{{\cal F}}
\newcommand{\pasl}{\pa\kern-.55em /}

\newcommand{\ksl}{k\kern-.55em /}

\newcommand{\tQ}{{\tilde Q}}
\newcommand{\tq}{{\tilde q}}
\newcommand{\tT}{{\tilde T}}
\newcommand{\bj}{{\bar j}}

\DeclareFixedFont{\xiiss}{OT1}{cmss}{m}{n}{12}
\DeclareFixedFont{\ixss}{OT1}{cmss}{m}{n}{9}
\DeclareFixedFont{\cmrnine}{OT1}{cmr}{m}{n}{9}
\newcommand{\field}[1]{\mathbb{#1}}
\newcommand{\BC}{{\field C}}

\newcommand{\BZ}{{\field Z}}

\newcommand{\CCs}{\hbox{\ixss C\kern-.4emI}}
\newcommand{\ZZs}{\hbox{\ixss Z\kern-.4emZ}}

\newcommand{\CA}{{\cal A}}

\newcommand{\CI}{{\cal I}}
\newcommand{\CM}{{\cal M}}
\newcommand{\CN}{{\cal N}}
\newcommand{\CP}{{\cal P}}

\def\toricrefs{\cite{BP,CFIKV,FHHU}}

\title{Seiberg Duality for Quiver Gauge Theories}

\author{David Berenstein$^\dag$ and Michael R. Douglas $^\ddag$\\
$^\dag$ School of Natural Sciences, 
Institute for Advanced Study, 
Princeton, NJ 08540, USA\\
$^\ddag$ Department of Physics and Astronomy,
Rutgers University,
Piscataway, NJ 08855-0849, USA\\
$^\ddag$ Isaac Newton Institute for Mathematical Sciences,
Cambridge, CB3 0EH, U.K.
\\
$^\ddag$
I.H.E.S., Le Bois-Marie, Bures-sur-Yvette, 91440 France
}

\abstract{
A popular way to study $\CN=1$ supersymmetric gauge theories is to
realize them geometrically in string theory, as suspended brane
constructions, D-branes wrapping cycles in Calabi-Yau manifolds,
orbifolds, and otherwise.  Among the applications of this idea are
simple derivations and generalizations of Seiberg duality for the
theories which can be so realized.

We abstract from these arguments the idea that Seiberg duality arises
because a configuration of gauge theory can be realized as a bound
state of a collection of branes in more than one way, and we show that
different brane world-volume theories obtained this way have matching
moduli spaces, the primary test of Seiberg duality.

Furthermore, we do this by defining ``brane'' and all the other
ingredients of such arguments purely algebraically, for a very large
class of $\CN=1$ quiver supersymmetric gauge theories, making physical
intuitions about brane-antibrane systems and tachyon condensation
precise using the tools of homological algebra.

These techniques allow us to compute the spectrum and superpotential
of the dual theory from first principles, and to make contact with
geometry and topological string theory when this is appropriate, but
in general provide a more abstract notion of ``noncommutative
geometry'' which is better suited to these problems.  This makes contact
with mathematical results in the representation theory of algebras;
in this language, Seiberg duality is a tilting equivalence between
the derived categories of the quiver algebras of the dual theories.
}

\preprint{RUNHETC-2002-26}

\begin{document}

\section{Introduction}

One of the more striking results in the modern study of $\CN=1$
supersymmetric gauge theory was Seiberg's discovery \cite{Seiberg}\ of
an IR duality between two QCD-like theories, both with $N_f$ flavors
of quarks (fundamental chiral superfields), but with different gauge
groups $SU(N_c)$ and $SU(N_f-N_c)$.  The simplest prediction of this
duality is that the moduli spaces of supersymmetric vacua of the two
theories are the same.  Other nontrivial agreements between the
theories are the 't Hooft anomaly matching conditions, and the
behavior under adding relevant operators.  This duality was extended
to many $\CN=1$ theories, and various physical derivations of it were
proposed, as reviewed in \cite{ILS, BS, GK}

Perhaps the simplest and most suggestive derivation was in terms of
the Hanany-Witten-Diaconescu suspended brane construction 
\cite{HW,Diacmono}.  As
in all such constructions, one obtains $D$-dimensional gauge theory as
the world-volume theory of a set of Dirichlet branes extending in
$D$ dimensions; the embedding in the additional or ``internal'' $10-D$
dimensions determines the spectrum and other structure.  Here, the
D-branes are strings in the internal dimensions, with one or both ends
attached to NS$5$-branes.  Taking $N_c$ finite length strings between
$5$-branes produces pure $SU(N_c)$ gauge theory, and adding $N_f$
semi-infinite strings produces $N_f$ quarks.  The duality is then
obtained by moving one $5$-brane around the other to exchange their
positions, which reverses the orientation of the finite strings.  If
$N_f\ge N_c$, by moving and reconnecting the strings, one obtains a
theory with $N_f$ semi-infinite and $N_f-N_c$ finite strings, the
dual theory. \cite{EGK}

This construction provides a very intuitive physical picture for the
duality and shows clearly where it is relevant in compactification of
string theory: since the inverse coupling constant is identified with
the length of the string, it will arise if one varies the gauge coupling
(by varying a moduli field) through infinity.  The result of
this is to turn branes into their antibranes, leading to a very simple
derivation of the flip $N_c \rightarrow -N_c$.

The main disadvantage of this construction is that certain steps are
somewhat {\it ad hoc}.  In particular, reproducing the moduli spaces
of vacua requires postulating rules such as the ``s-rule'' of \cite{HW} on
allowed brane configurations.  Although this particular rule can be
justified independently \cite{GK}, one is left with a qualitative approach
which works only in simple examples.  One requires more precision to
treat more complicated examples, and ultimately only a procedure which
can produce explicit superpotentials can be considered completely
satisfactory.

Of course gauge theories can be embedded in string and M theory in
many ways.  A large class of $\CN=1$ examples is provided by the quiver
gauge theories of D-branes at $\BC^3/\Gamma$ orbifold singularities
\cite{DGM}.  This approach is both highly computable, and has a direct
geometric interpretation in terms of branes wrapping cycles in the
internal space obtained by resolving the singularity.

Recently, several groups \toricrefs\ have shown that Seiberg
dual pairs or even sets of theories can be obtained in this framework,
using the procedure of ``partial resolution'' \cite{morpless}.  Many
interesting singularities are not orbifold singularities, but can be
obtained by partial resolution of an orbifold singularity.
Physically, the gauge theory associated to the singularity is obtained
by starting with the orbifold quiver theory, turning on certain
Fayet-Iliopoulos terms, and integrating out any matter which becomes
massive.  One can make various choices of FI terms and their signs,
and these choices lead to Seiberg dual theories.

The intuitive picture is very similar to the work on branes in flat
space.  As before, one can take a gauge coupling through infinity
through a continuous deformation which changes the sign of FI terms.
This leads to a topologically distinct but birationally equivalent
resolution.  The identification of ``fractional brane'' to brane
wrapping cycle depends on this choice, so one gets different geometric
identifications related by ``brane-antibrane'' transitions of the sort
described above.  This work is thus one part of the motivation for the
present work.

In fact, one does not need to bring in the geometry of orbifold
resolution in this discussion, and one can discuss Seiberg duality for
general quiver theories, not just those obtained from orbifolds.  A
more direct motivation for the present work is an observation of
B. Fiol, that Seiberg duality is related to the ``reflection functor''
of the theory of quiver representations.  We will show that this is
indeed what underlies classical duality in Seiberg's original example,
and provides a one-to-one map between configurations.

In simple cases, this map can be understood quite simply as
translating a bound state formed from a particular basis $B_i$ of
branes, to a new but simply related basis $B'_i$, copying over all the
structure of the bound state.  For example, the suspended brane
argument is the change of basis from $(B_1,B_2)$ the semi-infinite and
finite strings, to $(B_1+B_2,-B_2)$ a new basis of a semi-infinite
string obtained by combining $B_1$ and $B_2$, with the
orientation reversed finite strings $B_2$.

However, our map is defined in a completely algebraic way, explicitly
computing the superpotential of the dual theory, and mapping specific
field configurations to field configurations, without requiring other
{\it ad hoc} input.  For purely bifundamental matter, the relation
between superpotentials is indeed the one postulated by
\cite{BP,CFIKV,FHHU}.  However, our argument generalizes to a much
larger class of quiver theories, and we will demonstrate this by
treating the adjoint theories of \cite{KSS}.  We feel our arguments
both explain the duality concretely, and reduces the general problem
of finding dualities to mathematics, in a sense we will describe.

As in the suspended brane arguments, our basic results will be for
$U(N)$ gauge theories with large Fayet-Iliopoulos terms.  This
justifies treating the theories classically, and our basic results
will be to match the moduli space of pairs of classical theories.
We will also compute the superpotential for the dual theory in terms
of the original theory.  Again this will be for large FI terms, but
if we make the natural assumption (which can be proven in some cases)
that the superpotential is independent of the FI terms, then this is
the general result.  In this sense we can say we have proven Seiberg
duality for this class of theories.

Of course one cannot avoid discussing quantum corrections in general.
We will not much discuss the quantum aspects, not because we think this
is unimportant or because they cannot be addressed, but more because at
the moment we only see how to do this on a case by case basis, as is
already well discussed in the literature.  

To summarize what follows, in section 2 we give a very explicit
argument for duality in the original case of supersymmetric QCD (with
$U(N)$ gauge group), to make the ideas clear.  We will be able to make
almost all of the argument in purely field theoretic terms, without
explicitly bringing in string theory.  In section 3 we generalize this
to arbitrary $U(N)$ quiver theories with a superpotential.

In section 4 we explain how many of these arguments can be based on the
theory of Dirichlet branes in Calabi-Yau manifolds.  This will allow
us to clean up some points in the previous argument and serve as a
simple introduction to the derived category, which underlies the
discussion.  We then reconsider our examples and move on to treat
the case with an adjoint field.

In section 5 we discuss related mathematics.  In particular, there is
a theorem \cite{Rickard} which, in a sense we will explain, shows that
all Seiberg-like dualities are examples of ``tilting equivalences,'' a
generalization of the reflection functor.  This will allow us to give
a simple general argument that toric dualities are indeed Seiberg
dualities.

We also go somewhat more deeply into the underlying formalism, to make
the point that the homological algebra we are using does not depend
{\it a priori} on realizing the gauge theories using branes in a
Calabi-Yau, but can be defined directly, just from the quiver gauge
theory.  

Finally we give conclusions in section 6.

\section{Seiberg duality as change of basis}\label{sec:chabas}

In this section we consider the original example of $SU(N_c)$ gauge
theory with $N_f$ flavors of massless quarks, i.e. $N_f$ chiral
multiplets in the fundamental $N_c$ and $\bar N_c$ representations.
Classically,
this theory has global symmetry $U(N_f) \times U(N_f)$ acting on the
$N_c$ and $\bar N_c$ quarks respectively.

Our arguments will essentially be classical, so we ignore anomalies
for the time being.  While we are at it, we can even take the numbers
of $N_c$ and $\bar N_c$ quarks to differ.  Furthermore, we ignore for
now the difference between $SU(N)$ and $U(N)$, and take $U(N)$ gauge
group.

\begin{figure}[ht]
\begin{center}
\leavevmode
\epsfxsize=8cm
\epsfbox{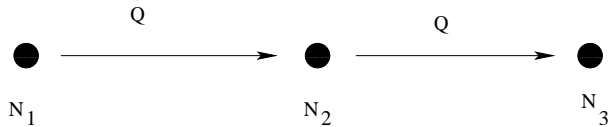}
\end{center}
\caption{Quiver diagram corresponding to the $(N_1,N_2,N_3)$ theory.}
\label{fig:quiver1.eps}
\end{figure}

We write the resulting continuous symmetry group as
\begin{equation}\label{eq:theoryone}
U(N_1) \times U(N_2) \times U(N_3) ,
\end{equation}
and the matter is chiral multiplets $\tQ$ and $Q$ transforming as
\begin{equation}\label{eq:spectrumone}
\tQ \in (\bar N_1, N_2, 1);  \qquad Q \in (1,\bar N_2,N_3) .
\end{equation}
This theory can be summarized in the diagram in figure
\ref{fig:quiver1.eps}.

Clearly these theories can be obtained by embedding or wrapping
Dirichlet branes in a variety of ways.  The standard (non-chiral) theory
can be obtained from any configuration containing two
distinct types of brane, $B_c$ and $B_f$, such that the
massless fermionic open strings between $B_c$ and $B_f$ are precisely one
left and one right chirality Weyl fermion.  The generalized theories
can be obtained by postulating branes $B_1$, $B_2$ and $B_3$, with
a single Weyl fermion between the pair $(B_1,B_2)$ and between
the pair $(B_2,B_3)$.
The spectrum, general
properties of D-branes and supersymmetry lead directly to the
classical quiver theory (diagram).
We will sometimes denote the theory with brane content
$N_1 B_1 + N_2 B_2 + N_3 B_3$ as the theory
``$(N_1\ N_2\ N_3)$'', and refer to this vector as the
``charge.''

We will only consider quiver theories, with bifundamental and adjoint
matter, for which the notation \rfe{spectrumone}\ is somewhat
cumbersome.  We now switch to denote the vector space of chiral
multiplets charged
$$
(\ldots,\bar N_E,\ldots,N_{E'},\ldots)
$$
for two branes $E$ and $E'$ as
$$
\Ext(E,E') .
$$
In other words, a theory with $k$ such chiral multiplets would have
$\Ext(E,E')=\BC^k$, and the consequent $U(k)$ global symmetry (which
could of course be broken by the superpotential) would act linearly
on this vector space.
A similar notation to denote the
complex gauge invariances or gauginos is
$$
\Hom(E,E') .
$$
Normally one has $\dim\Hom(E,E') = \delta_{E,E'}$ for a collection
of branes each carrying a $U(1)$ world-volume gauge field.

In this notation, $\tQ\in\Ext(B_1,B_2)$ and $Q\in\Ext(B_2,B_3)$.
At this point, this is just an alternate way to describe the field
content of the world-volume theory.  Once we define bound states of
branes, we will use the same notations to denote the linear spaces of
chiral multiplets and gauge symmetries between any pair of branes.

\subsection{BPS bound states and duality as change of basis}

We now ask, what are the simple BPS bound states of branes in this theory,
meaning combinations of branes, possibly with vevs of chiral fields,
which preserve supersymmetry and break the gauge symmetry to $U(1)$.

It is fairly obvious for the theory \rfe{theoryone}, and can even be
proven rigorously, that there are three such bound states,
with charges
\begin{equation}\label{eq:boundstates}
[B_4] = (1\ 1\ 0); [B_5] = (0\ 1\ 1); [B_6] = (1\ 1\ 1) .
\end{equation}
These supersymmetric gauge theories have one (for $B_4$ and $B_5$) or
two (for $B_6$) chiral multiplets, which must be set non-zero to break
the gauge symmetry to $U(1)$.  Their precise values, up to gauge
equivalence, are completely determined by the FI terms, which must be
set non-zero (in an obvious way) for any of these to be supersymmetric
vacua.  Thus, all three bound states are ``rigid,'' meaning that they
themselves have no moduli, and are described at low energy by pure
supersymmetric $U(1)$ gauge theory.

We can ignore the dependence on FI terms for now by defining a
``holomorphic configuration'' of the gauge theory as a configuration
of the chiral multiplets up to complexified gauge equivalence, here
with complex gauge group $\prod_i GL(N_i,\BC)$.  Thus we define (for
example) the brane $B_4$ as the unique holomorphic configuration of
$(1\ 1\ 0)$ quiver gauge theory with unbroken $GL(1)$ gauge symmetry.
This is the precise sense in which we define a brane $B_i$ as a bound
state of branes (but not antibranes as yet).  We will then use the
notation $[B_i]$ to denote the charge of the brane $B_i$.

More generally, a theory with given charge might have more than one
gauge equivalence class of supersymmetric configuration.  In general
we will use the notation $B_i$ for a brane, to refer to a particular
holomorphic configuration of the theory with particular charges.

The main idea is now that, since the bound states \rfe{boundstates}\ %
share all of the properties of the original branes $B_1$, $B_2$ and
$B_3$, it should be possible to describe any configuration made out of
the original branes equally well as a bound state of the new branes,
or as a bound state of some combination of the old and new branes.
Each such choice of a generating set of branes will lead to an {\it a
priori} different supersymmetric gauge theory.  Since we are merely
reexpressing the same brane configuration as a bound state in a
different way, if in each theory there is a single way to get each
configuration as a bound state (this is what we will mean by a set of
branes forming a ``basis''), this must give a one-to-one map between
gauge theory configurations.  This idea can be made quite precise and
will lead to our map.

Suppose we try to go from the basis $(B_1,B_2,B_3)$ to the basis
$(B_1,B_2,B_5)$.  Following the intuition that the ranks of the gauge
groups are charges, and inverting to get $[B_3] = [B_5] - [B_2]$,
we can identify the new charges $N'_i$ in terms of
the old $N_i$ as
\begin{equation}\label{eq:dualityone}
N'_1 = N_1;
N'_5 = N_3;
N'_2 = N_2 - N_3 .
\end{equation}
Evidently there are two cases, $N_2 \ge N_3$ and $N_2 \le N_3$.
We can try to proceed either way, but in the second case $N_2 \le N_3$,
we should instead of $B_2$ use its antibrane $\bar B_2$, with multiplicity
$$
N'_{\bar 2} = N_3 - N_2 .
$$

\subsection{Dual world-volume theory}

The next step is to derive the world-volume theory of a collection of
branes from the new basis.  Since these are bound states of the
original branes, everything should be determined by straightforward
computation.

The idea behind this is that the massless fields associated to a pair
of branes $E$ and $E'$ with charges $N_i$ and $N'_i$, each a specific
bound state, can be found by computation in the joint $\prod_i
U(N_i+N'_i)$ theory describing the pair.  For the case with no
superpotential, this is described in some detail
in the appendix to \cite{DFR}, and we give the basic idea here.
A more general discussion covering some superpotentials is in
\cite{Trieste}.

Consider the
combinations of $B_5$ with $B_1$ and $B_2$ relevant for the problem at
hand.
We first combine $B_1$ and $B_5$.  Together, these sit in the gauge
theory $(1\ 1\ 1)$.  Upon combining this pair of branes, there is an
extra massless field, inherited from that between $B_1$ and $B_2$,
and transforming under the unbroken gauge groups of $B_1$ and $B_5$.
This can be summarized in the result
\begin{equation}\label{eq:extdimone}
\dim \Ext(B_1,B_5) = 1 .
\end{equation}

For $B_2$ and $B_5$, one must consider the theory $(0\ 2\ 1)$.
Upon bringing this pair together, we also gain
an extra matter field, now from $\Ext(B_2,B_3)$.
However, unlike the first case, the vev of this field
can be gauged away by $GL(2)$, so it does not contribute a physical
degree of freedom.  In fact there are no extra massless
chiral multiplets:
\begin{equation}
\dim\Ext(B_2,B_5)=\dim\Ext(B_5,B_2)=0 .
\end{equation}

An equivalent way to say this is to note that the combined theory has
two extra vector multiplets, off-diagonal in $GL(2)$.  If the vev of
the $\Ext(B_2,B_3)$ chiral multiplet had been zero, i.e. $Q=0$, the
$GL(2)$ would be unbroken, and these would transform as a
$\Hom(B_2,B_5)$ and a $\Hom(B_5,B_2)$.  On the other hand, in the
bound state $B_5$, $Q\ne 0$.  This vev Higgses the $\Hom(B_2,B_5)$
gauge boson, and both chiral and vector multiplet are lifted. It is
Higgsed by the matter, and both become massive.
However, the $\Hom(B_5,B_2)$ vector remains massless.

The resulting massless spectrum is
\begin{equation}\label{eq:homdimone}
\dim \Hom(B_5,B_2) = 1
\end{equation}
with all other dimensions zero.

This is the basic data we now need to incorporate in a new, presumably
dual, quiver gauge theory.  We have one node for each of $B_1$, $B_2$
and $B_5$.  The result \rfe{extdimone}\ is easy to incorporate as this
tells us that we have a single bifundamental chiral multiplet
in the $(\bar N'_1,N'_5)$.  Encouragingly, this has the right quantum
numbers to be the ``meson'' field $M$ of \cite{Seiberg}.

On the other hand, the result \rfe{homdimone}\ is confusing, for
various reasons.  It tells us that when we put together two different
branes, $B_2$ and $B_5$, we get an enhanced gauge symmetry.  On the
other hand, one knows one does not get enhanced gauge symmetry when
one puts two different D-branes together.

Furthermore, the asymmetry between $B_2$ and $B_5$ leads to a
non-unitary gauge group, which might be disturbing.  There are various
reasons why this does not contradict general theorems, but the most
pertinent is that this configuration cannot solve the D-flatness
conditions (for any choice of FI terms) and thus cannot be realized as
a vacuum. \footnote{A related observation is that, in the physical theory,
a vev for the field $Q$ implies a vev for
its hermitian conjugate $Q^\dagger$.
This will give a mass to the other off-diagonal vector particle 
and lead to a unitary gauge group.  In our holomorphic definitions,
we are only giving a vev to $Q$.}
In the brane language, $B_2$ and $B_5$ cannot form a bound
state.

We will come back to this point later, but indeed it is true that this
makes it difficult to describe this combination of three branes
directly as a supersymmetric field theory.  Of course, we know that
not all combinations of branes can be so described, for example
combinations of a brane with its own antibrane are problematic.

However, one {\bf can} describe the combination $(B_1,\bar B_2,B_5)$
with supersymmetric field theory.  To see this, one needs a definition
of the antibrane $\bar B_2$.  Usually this is explained in terms of
the string world-sheet, or in terms of embeddings of branes.  For
example, in the D-brane context, an antibrane is defined by reversing
the GSO projection.  This exchanges vector and chiral multiplets, and
will relate the $\Hom(B_5,B_2)$ of \rfe{homdimone}\ to a chiral
multiplet $\Ext(B_5,\bar B_2)$.  This has the right quantum numbers to
be one of the ``dual quarks'' of \cite{Seiberg}, so clearly this is
a step in the right direction.

In fact, one can define antibranes without any reference to string
theory or a higher dimensional embedding, by using the language of
homological algebra, as we will now explain.

\subsection{The reflection functor}

We now explain how bound states of $B_2$ and $B_3$ map in a one-to-one
way to bound states of $\bar B_2$ and $B_5$.
This construction, known as a ``reflection functor'' in the
theory of quiver representations \cite{BGP},
will underlie the map between configurations in the general case.

The change of basis $(B_2,B_3) \rightarrow (\bar B_2,B_5)$ does not
directly involve $B_1$, so fow now let us just consider
$U(N_2)\times U(N_3)$ gauge theory with a single chiral
field $B$ in the $(\bar N_2,N_3)$.  We assume $N_3\ge N_2$ and
let $N=N_3$.

The basic problem we face is to give a definition to the antibrane
$\bar B_2$, using only the most general features of string theory.
Now the defining property of our new basis is that $B_3$ can be
obtained as a bound state of $\bar B_2$ and $B_5$.  Given the result
\rfe{homdimone}, there is a well-motivated mathematical way to do
this: we realize $B_3$ as a complex.  This is a construction of
homological algebra in which we take a finite length
sequence of ``objects'' (branes
for us) $E_i$, $i$ integer, and postulate a series of linear maps
$d_i:E_i\rightarrow E_{i+1}$ between successive terms in the complex,
satisfying $d_{i+1}\cdot d_i=0$.  The standard notation for
such a complex is
\begin{equation}\label{eq:defcomplex}
\complex{E_{-m} \mapr^{d_{-m}} E_{-m+1} \mapr^{d_{-m+1}} \ldots
 \mapr^{d_{-1}} {\ul E_0} \mapr^{d_0} E_1 \mapr^{d_1} \ldots
 \mapr^{d_{n-1}} E_n .}
\end{equation}
The underline is used to indicate the zero position in the complex.

At least in a first approximation,
we will interpret the {\it cohomology} of $d$ as the physical brane
represented by the complex.  The most elementary example of this is
a complex of the form
\begin{equation}\label{eq:vacuum}
\complex{ B \mapr^1 B }
\end{equation}
where the two terms are the same object $B$, and $d=1$.  We could also
consider the direct sum of this with another complex.  In either case,
this combination completely cancels out of the cohomology.  We
will use this to represent the physical fact that such a $B\bar B$ pair
can annihilate to the vacuum via tachyon condensation \cite{Sen}, 
and that the vacuum is the unique
supersymmetric configuration of this type.

To define this cohomology in more complicated problems, one of course
needs some definition of the underlying objects.  Here our definition
is holomorphic configurations of a quiver gauge theory, as in
\cite{DFR}.  For D-branes on Calabi-Yau, one might use coherent
sheaves on the Calabi-Yau as the underlying objects instead.  We will
discuss this point further in sections 4 and 5.  Actually, much of the
discussion can be made without knowing what the objects are, only that
they satisfy the axioms of an abelian category.  If this is too
abstract for the reader's taste, he is encouraged to ignore it.

To represent $B_3$ as a brane-antibrane bound state, we will use a two
term complex with $E_0=B_5$ and $E_1=B_2$, and $d =\alpha$ a non-zero
element of $\Hom(B_5,B_2)$.  We denote this as
$$
B_3 = \bracomplex{ {\ul B_5} \mapr^\alpha B_2 } .
$$
The cohomology does not depend on the overall magnitude of $d$ and
since in our example $\dim\Hom(B_5,B_2)=1$ there is a unique such
bound state.  More generally one gets a bound state with
$\dim\Hom(B_5,B_2)-1$ parameters or world-volume chiral multiplets.

There are various ways to interpret this construction physically.  One
is that $d$ represents a brane-antibrane tachyon, and we are
constructing the bound state $B_3$ by tachyon condensation.  Another
interpretation is that $d$ forms part of the BRST operator in
topological open string theory, as in \cite{DougDC}.  The two interpretations
are closely related in the case of a two-term complex, but the BRST
interpretation also makes sense for the case of longer complexes.

For now, we postpone these subtleties and will not rely on the
detailed physics of the construction, but use only its most basic
property: namely, that the complex \rfe{vacuum}\ is equivalent to the
vacuum configuration.

Having this definition of the change of basis, we now
seek a map which takes a configuration of this theory, in
other words a vacuum expectation value for $Q$ up to complex gauge
equivalence, and produces a unique corresponding
configuration of $U(N_3-N_2) \times U(N_3)$ gauge theory with a single
chiral field in the $(N_3-N_2, \bar N_3)$, call this $\tilde q$.

The starting configuration can now be realized as the complex
\begin{equation}\label{eq:Bcomplex}
\xymatrix{
\BC^{N_2} \otimes B_2 \ar[dr]^Q & & \\
\BC^N \otimes B_5 \ar[r]^{{\bf 1} \otimes \alpha\quad }  &\ \ \  \BC^N
\otimes B_2 }.
\end{equation}
Here ${\bf 1}_N$ is the $N\times N$ identity matrix, and the
vertical axis is just direct sum (e.g.,
the map $d$ is the direct sum of the maps indicated by both arrows).

This complex is obviously not the minimal complex we could use to
describe this state; a complex containing fewer branes would be
obtained by annihilating $B_2\bar B_2$ pairs.
If we assume that $Q$ has maximal rank (we discuss this below),
we can apply a $GL(N)$ transformation $g$ on the left,
\begin{equation}\label{eq:gaugetrans}
g \left(Q\ {\bf 1}_N\right) =
\left({\bf 1}_{N_2}\ g \right) =
\begin{pmatrix} {\bf 1}_{N_2}& v \\
      0& \tilde q
\end{pmatrix}
\end{equation}
to obtain a complex with $N_2$ $B_2\bar B_2$ pairs of the form \rfe{vacuum},
an $N\times (N-N_2)$ matrix $\tilde q$, and an
an $N\times N_2$ matrix $v$.
We then annihilate the $B_2\bar B_2$ pairs,\footnote{
Among the subtleties in defining this, one is particularly noteworthy.
Part of $d$ maps from the $B_5$'s to the $B_2$'s which will be
annihilated (this is $v\ne 0$), and in doing the annihilation, we
are losing this information.
Physically, this makes the annihilation process like an RG flow
(as indeed it is in boundary CFT).
Mathematically, it is formalized in the
statement that the equivalence between the two configurations is a
{\it quasi-isomorphism}, a chain map which preserves homology.
}
to obtain an equivalent complex
$$
\BC^N \otimes B_5
\mapr^{\tilde q} \BC^{N-N_2} \otimes B_2 .
$$
This is the representation of the same configuration in the ``dual''
theory.

One should ask whether the step of making a $GL(N)$ transformation
leads to some ambiguity in this procedure.  To see that
this procedure takes a gauge equivalence class of configurations of $Q$
to a single gauge equivalence class of configurations $\tilde q$,
we can rephrase the map in the following equivalent way.
We add another arrow to \rfe{Bcomplex},
$$
\xymatrix{
\BC^{N_2} \otimes B_2
\ar[d]^Q  \ar[dr]^Q & \\
\BC^N \otimes B_5  \ar[r]^{{\bf 1} \otimes \alpha} & \BC^N \otimes B_2 , }
$$
to get a commutative diagram (notice that we are allowed to do this
because the map $1\otimes\alpha$ is onto).  This property will of course
be preserved by the transformation \rfe{gaugetrans}.
Annihilating the $B_2\bar B_2$ pairs then restricts the product
${\tilde q} \cdot Q$ to
an ${N-N_2}$-dimensional subspace, the quotient space
$\BC^N / \im Q$.
By definition however this is
the cokernel of $Q$,  so one has the gauge invariant statement that
\begin{equation}\label{eq:deftildeq}
{\tilde q} \cdot Q = 0 .
\end{equation}
This can be regarded as a set of linear equations which given $Q$
determines $\tilde q$ up to gauge equivalence.
One can say the same thing by writing the exact sequence
\begin{equation}\label{eq:tildeBseq}
0 \mapr \BC^{N_2} \mapr^Q \BC^{N_3} \mapr^{\tilde q} \BC^{N_3-N_2} \mapr 0 .
\end{equation}

All this assumed that $Q$ has maximal rank.  On the other hand, if $Q$
did not have maximal rank, we would not be able to find a $GL(N)$
transformation $g$ of the form \rfe{gaugetrans}.  This does not
prevent us from making the change of basis, but doing it results in a
complex containing terms $B_2 \mapr^0 B_2$ which are ``non-annihilated
brane-antibrane pairs.''

Brane-antibrane annihilation can be defined more precisely by
using the formalism of the derived category.  In the derived category
it will be true that the map between configurations is one-to-one in
all cases, but will not always take configurations of supersymmetric
gauge theory to supersymmetric gauge theory.
The configurations with less than maximal rank typically (not always)
leave unbroken nonabelian gauge symmetry, and thus these cases obtain
non-trivial quantum corrections.  We will return to this later.

We can go on to derive the value of $M$ in the dual theory for
a given starting configuration, by putting in $B_1$ and its map
to the other branes.  Adding this to the complex \rfe{Bcomplex}\ %
produces the commutative diagram
$$
\xymatrix{
B_1 \otimes \BC^{N_1} \ar[r]^{\tQ\ }\ar[dr]^M & B_2 \otimes \BC^{N_2}\ar[dr]^Q \\
& B_5 \otimes \BC^{N_3} \ar[r]^{1\ } & B_2 \otimes \BC^{N_3},
}
$$
so clearly
\begin{equation}\label{eq:Mdef}
M= Q\cdot \tilde Q .
\end{equation}
This is gauge equivalent using \rfe{gaugetrans}\ to
$$
\xymatrix{
B_1 \otimes \BC^{N_1}\ar[dr]^M \ar[r]^\tQ& B_2 \otimes \BC^{N_2}\ar[dr]^1 \\
& B_5 \otimes \BC^{N_3} \ar[r]^g & B_2 \otimes \BC^{N_3}.
}
$$
The same argument which led to \rfe{deftildeq}\ then leads to
\begin{equation}\label{eq:relone}
\tilde q \cdot M = 0 .
\end{equation}

Given the constraints that $Q$ and $\tilde q$ are maximal rank,
the map from $(\tilde Q,Q)$ to $(M,\tilde q)$ is one to one.
In the forward direction this is clear.  In the reverse direction,
this follows by reinterpreting \rfe{deftildeq}\ as a set of linear
equations determining $Q$ from $\tq$.

There is a very similar transformation
which maps configurations of $B_1$ and $B_2$ to configurations of
$B_4$ and $\bar B_2$.  The analysis can be reduced to the previous
case by the device of reversing all the arrows, and leads to a map
from $\tilde Q$ to $q$ defined by
\begin{equation}\label{eq:tildeAseq}
0 \mapr \BC^{N_1-N_2} \mapr^{q} \BC^{N_1} \mapr^{\tilde Q} \BC^{N_2}
 \mapr 0 .
\end{equation}

\subsection{The role of the FI terms}

Continuing the discussion of the change of basis to $(B_1,\bar B_2,B_5)$,
we next need to derive the massless spectrum between $B_1$ and
$\bar B_2$, i.e. the other dual quark.  Again there are string
theory arguments for this, which we will return to, but again
we would like to see if this can be done purely algebraically.

Naively, one might think we could just repeat what we did to get the
first dual quark, as the final dual theory is symmetric under exchanging
their roles.  In fact this is completely wrong.  Our procedure treats
the two very asymmetrically, because we replaced $B_3$ with $B_5$, keeping
$B_1$ in the new basis.  We could have made the alternate choice of
keeping $B_3$ and replacing $B_1$ with $B_4$, but this is a different
change of basis.

In fact this asymmetry is inherent in the problem, in the classical
limit in which we are working.  This limit is best justified by the
device of turning on large Fayet-Iliopoulos terms, which completely
break the gauge symmetry at high energy.  This makes quantum effects
arbitrarily small, and in fact the exact moduli space agrees with the
classical result.  However, in turning on FI terms, one is making a
choice which leads to the asymmetry.  Indeed, varying FI terms will
lead to naturally to variations of the basis.

The D-flatness conditions for SQCD are
$$
\tQ^\dagger \tQ - Q Q^\dagger = \zeta \cdot {\bf 1}
$$
where $\zeta$ is the FI parameter.  If we take this large and positive,
the vacuum expectation value of $\tQ$ is forced to take its maximal
rank $N_c$.  There is a vacuum with $Q=0$;
using $U(N_f)$ symmetry it can be brought to the form
$$
\tQ = \left(\sqrt{\zeta} \cdot {\bf 1}_{N_2}\ \  {\bf 0}_{N-N_2}\right).
$$
Giving a vev to $Q$ increases the magnitude of $\tQ$, which always has
maximal rank.

Conversely, if $\zeta$ had been large and negative, we would find
supersymmetric vacua in which $Q$ always had maximal rank.

The reflection functor is a one-to-one map between gauge theory
configurations precisely when the morphisms involved have maximal
rank, so we now see that we can obtain a duality in either
of these limits, but by using two different changes of basis.

The change of basis to $(B_1,\bar B_2,B_5)$ is appropriate when
$\zeta>>0$.  The relations \rfe{tildeBseq}\ and \rfe{Mdef}\ give us
a map from $(\tilde Q,Q)$ to $(M,\tilde q)$, satisfying \rfe{relone},
\begin{equation}
\tilde q \cdot M = 0 .
\end{equation}

The existence of these vacua tells us that the dual theory also has
large FI terms.
These nonzero FI parameters are also implicit in brane treatments, as
we discuss in the next section.  In this context one can independently vary 
the FI parameters for each node.  The original theory then
corresponds to the regime
\begin{equation}\label{eq:origFI}
\zeta_1 < \zeta_2 < \zeta_3
\end{equation}
while the dual we discussed corresponds to
\begin{equation}\label{eq:dualFI}
\zeta_1 < \zeta_5 < \zeta_2 .
\end{equation}

\subsection{Superpotential and completing the argument}

To interpret these as fields in a dual gauge theory, the relation
\rfe{relone}\ must follow from the classical equations of motion or other
constraints in this theory.  There is only one way to accomplish this,
which is as an F-flatness condition.\footnote{
The relation lifts fermions, so it cannot be a D-flatness condition.
}
Thus we postulate a
superpotential $W$ and an additional chiral field $q$ in
$\Ext(\bar B_2,B_1)$ such that
\begin{equation}
\frac{\p W}{\p q} = 0
\end{equation}
implies \rfe{relone}.  The simplest candidate is
\begin{equation}\label{eq:superone}
W = \tr M q \tilde q .
\end{equation}

This step may seem less well motivated than the previous ones, but in
fact making it requires more input than just algebra.  We need to
assume that the F-flatness conditions (quiver relations) in the dual
theory follow from a superpotential, or else a statement from which
this follows.  In section 4, we will derive the existence of the field
$q$ and the superpotential \rfe{superone}\ in the weakly coupled
D-brane context.

One can write more complicated superpotentials which lead to
\rfe{relone}.  In fact, we suspect that general arguments should not
produce a unique answer unless we specify additional conditions or
give a particular UV definition of the theory.  The simplest
additional condition which determines \rfe{superone}\ is to insist that
the spaces of supersymmetric vacua be the same on both sides.  If we
maintain the field content we discussed, the general gauge-invariant
superpotential is $W = \tr f(M\tq q)$ for a holomorphic function $f$.
Adding higher order terms to this always leads to additional vacua.
The overall coefficient of the linear term can of course be absorbed
into the normalization of the dual quarks.

Finally,
given \rfe{superone}, we have the additional F-flatness conditions
\begin{equation}
M q = q \tilde q = 0 .
\end{equation}
We can solve these and still have a one-to-one map by postulating
\begin{equation}\label{eq:zeroq}
q = 0 .
\end{equation}
This is also natural given the FI terms \rfe{dualFI}.

Thus the matter content of the dual theory is reproduced exactly.

\begin{figure}[ht]
\begin{center}
\leavevmode
\epsfxsize=7cm
\epsfbox{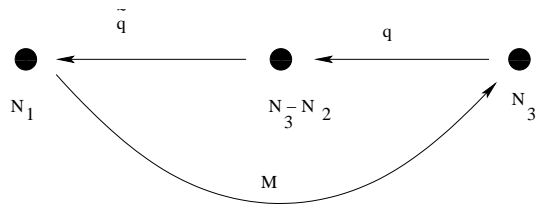}
\end{center}
\caption{Dual quiver diagram for theory $(N_1,N_2,N_3)$.}
\label{fig:quiver2.eps}
\end{figure}

To summarize the results: the change of basis
to $(B_1,\bar B_2,B_5)$ leads to a new gauge theory with quiver
as in \ref{fig:quiver2.eps}.
Moreover, when the dual ranks \rfe{dualityone} are non-negative,
there is a one-to-one map between supersymmetric
configurations (with large FI term)
given by \rfe{tildeBseq}, \rfe{Mdef}\ and \rfe{zeroq}.

\subsection{Other aspects of the duality}

The argument above may have struck the reader as long-winded,
especially as it only led to a subset of the original results of
\cite{Seiberg}.  However, the concepts we just defined allow
generalizing the same argument to a large class of quiver theories.

We will do this in the next section, but let us discuss some other
aspects of Seiberg duality in this example, which one might also try
to generalize.  Our point is not to make a strong claim that our
classical discussion can ``prove'' quantum Seiberg duality, but to try
to see how much can be captured at this level, and by general
arguments.

The main point of \cite{Seiberg} was of course that in a pair of dual
theories, one theory would be preferred because it gave a weakly
coupled description.  The direct physical implications of this
for $\CN=1$ superconformal field theories are not so obvious.  On the
other hand, one can add small mass terms to get more conventional
theories with a particle spectrum, and then the two different gauge
groups and sets of chiral fields would correspond to two candidate
particle spectrums.  Deciding which is preferred in general requires
knowing the low energy gauge couplings.  There are general results
on this for quiver theories \cite{CFIKV,Fiol}, but this is somewhat
tangential to our concerns here.

Other tests of the duality include matching of anomalies and chiral
rings. 
For $U(N)$ quiver theories, anomaly matching for the explicit $U(N_i)$
symmetry groups is fairly simple, as almost all of it follows from the
simple condition that each node have the same number of fundamentals
minus antifundamentals before and after the duality.  On the other
hand, anomaly matching for other global symmetries, involving the
$U(1)_R$ symmetry, is not manifest.  There are general arguments which
prove anomaly matching for certain pairs of theories with the same IR
moduli space \cite{DotMan}.  On the other hand, in general ``naive''
anomaly matching can fail, because a symmetry visible on one side can
be an accidental symmetry of the IR limit not manifest in the
classical Lagrangian, or because of strong coupling corrections to
anomalous dimensions (for the $U(1)_R$ anomaly matching).  These
subtleties indeed appeared for the toric dual theories considered by
\cite{BP}.  Given that anomaly matching is not manifest, it is not
clear at present in what sense one should try to ``prove'' it.

The chiral ring by definition is the algebra defined by multiplying
operators modulo the F-flatness relations $W'=0$.  One can of course
make field redefinitions and so the equivalence of chiral rings should
be formulated in a more geometric way which allows for this.  We
believe that such an equivalence should generally follow from the
equivalence of classical moduli spaces (with source terms added for
the fields).  We will not try to study this here but only remark that
the simplest picture of the situation is the standard mathematical
philosophy which considers a space to be equivalent information to the
algebra of functions on that space.  Here, the space is the space of
supersymmetric configurations, and the algebra is the chiral ring.

Our general philosophy, that Seiberg duality is the description of the
same configuration using different bases of branes, suggests that one
should have duality in some sense for any ranks of the gauge and
symmetry groups, not just those which lead to superconformal theories.
Let us consider this point.

Our arguments did not assume $N_f\le 3N_c$, and work for any
$N_f>N_c$.  However we need to discuss the other cases.  For
$0<N_f<N_c$ there are no supersymmetric configurations.  At zero FI
term this is because of quantum corrections to the superpotential,
while at nonzero FI term it is because one cannot solve the D-flatness
conditions.  We would say that there is still a dual theory in this
case, but it is a theory with $N_c'=N_f-N_c<0$ which has no
supersymmetric vacua.

The case of $N_f=N_c$ is the trickiest as there are supersymmetric
configurations, whose quantum corrected moduli space depends
explicitly on the quantum scale $\Lambda$.  It would seem rather
artificial to try to reproduce this in classical terms, but let us
make one comment about this.

According to the definitions we are giving the dual is a $U(N_c)$
theory with $N_c=0$.  It is not clear what an FI term would mean for
such a theory, to say the least.  But it is here that one needs to
look, for the following reason.  The usual discussion is for $SU(N)$
theory.  We neglected this distinction (as is usually done in brane
arguments) because the $U(1)$ sector only has trivial dynamics.  One
still has extra baryonic operators in the $SU(N)$ theories.  However,
very generally, one can trade these invariants for the $U(N)$ FI
terms, as discussed in \cite{morpless,BP}.  

In some more general phrasing of the problem, this relation between
baryonic operators and FI terms might make sense for $N_c=0$ as well,
and describe the quantum corrections.  Such a phrasing might be more
natural in string theory, where the FI terms are themselves controlled
by dynamical fields.

\section{General case -- a physical approach}

The arguments we just gave generalize very straightforwardly to
arbitrary theories with bifundamental matter.  We start with the case
of no initial superpotential, and then incorporate a superpotential.

The ideas generalize directly to any theory of ``finite representation
type'' \cite{ARS,Gabriel,He}, {\it i.e.} a theory with finitely many
simple brane bound states.  In particular this includes theories with
adjoint matter, as long as there is a superpotential which lifts the
moduli space of simple branes.  We begin the discussion of SQCD with
one adjoint along these lines, but eventually explicit computations
along the previous lines become tedious, at which point we break off
to develop more formalism.

\subsection{The general reflection functor}

The reflection functor can be defined more generally, on a quiver with
a node $E$ (playing the role of $B_2$ above) with $k$ arrows leaving
$E$; let their targets be $F_i$ for $1\le i\le n\le k$, with $n_i$
arrows from $E$ terminating on $F_i$.  It can also be applied to a
subquiver of this form in a larger quiver.  Let the numbers of branes
involved be $N_E E + \sum_i N_i F_i$.

The change of basis is now
$$
E \rightarrow \bar E; \qquad
F_i \rightarrow \hat F_i \equiv F_i + n_i E .
$$
A more precise definition of the second line is given by the exact
sequence
\begin{equation}\label{eq:defFMT}
0 \mapr F_i \mapr \hat F_i \mapr^{f_i} \Ext(E,F_i) \otimes E \mapr 0
\end{equation}
where $f_i$ is the ``tautological $\Ext$,'' essentially the direct sum
over a basis for $\Ext(E,F_i)$.  This leads to the change of rank
$$
N_E \rightarrow \sum_i N_i n_i - N_E .
$$

Considerations similar to those above (and which will be made precise
in section 4) then lead to
$$
\dim \Ext(\hat F_i,E) = n_i
$$
and a quiver with the same form as before but with arrows reversed.

Let the chiral multiplets be $\phi_\alpha$;
the generalization of \rfe{tildeBseq}\ is then
\begin{equation}\label{eq:tildeEseq}
0 \mapr \BC^{N_E} \mapr^\phi \BC^{\sum N_i n_i}
  \mapr^{\tilde \phi} \BC^{\sum N_i n_i - N_E} \mapr 0 .
\end{equation}

This multibrane generalization can also be applied to a quiver
containing a node $E$ and a set of nodes $G_i$ with arrows from
$G_i$ to $E$, by reversing arrows in the previous argument.

This reflection functor allows us to analyze Seiberg duality in a
general theory with nodes $B_a$ and no superpotential $W$.  We pick
one node, call it $E$, with $J$ incoming and $I$ outgoing arrows, and
call the corresponding fields $\tQ^j$ and $Q^i$, where the indices run
$1\le j\le J$ and $1\le i\le I$.  We then just repeat the same
arguments ignoring the fact that different arrows may terminate on
different nodes $B_a$, to get a dual theory with $Q$ and $\tQ$
replaced by $q$, $\tq$ and $M$ and a cubic superpotential.

This leads to a quiver with nodes $\hat F_i$, $\bar E$,
and dual quarks obtained by ``reversing all the arrows'' from $E$
to $F_i$.  The arguments which led to \rfe{Mdef}\ also go through
straightforwardly to give an $I\times J$ matrix of mesons
$$
M^{i\bar j} = Q^i \cdot \tQ^{\bar j}
$$
satisfying
\begin{equation}\label{eq:newrel}
\sum_{\bar j} \tilde q_{\bar j} M^{i\bar j} = 0 .
\end{equation}
Continuing the line of reasoning above, we require dual quarks $q^i$
to enforce these relations, and we postulate the natural cubic
superpotential.

\subsection{An initial superpotential}

We next generalize this argument to deal with a
superpotential, say $W(\phi,Q,\tQ)$ where $\phi$ are all the fields
which are singlet under the $U(N_E)$ gauge group and not mentioned in
the previous discussion.

The shortest way to do this is to note that the original F-flatness
conditions pick out a subset of the original configurations,
defined by equations
\begin{eqnarray}\label{eq:orgrel}
0 &= \frac{\p W}{\p\phi_\alpha} = F_\alpha(\phi,M) \\
0 &= \frac{\p W}{\p Q^i} = \tQ^\bj G_{i\bj}(\phi,M) \\
0 &= \frac{\p W}{\p \tQ^\bj} = G_{i\bj}(\phi,M) Q^i ,
\end{eqnarray}
where
$$
G^{ij} = \frac{\p}{\p M_{ij}} W|_{M^{i\bj} = Q^i\tQ^{\bj}} .
$$
This would be the same as $\frac{\p^2 W}{\p Q^i \p\tQ^{\bj}}$ except that
we insist that the two matrices $Q$ and $\tQ$ appear in succession.

Let us first assume that $G^{ij}$ does not depend on $M_{ij}$, i.e. if
$W$ is linear in $Q$ and $\tQ$, and return later to the general case.
Given this assumption, we can argue separately
for each configuration of the fields $\phi$.  In any particular
configuration, $G^{ij}$ will take a definite value, which we now treat
as fixed.

There are two general cases.  If the matrix $G$ has rank greater
than $N_f-N_c$, no choice of $Q$ and $\tQ$ can lead to a supersymmetric
vacuum, because the number of remaining massless fermions is $N_f < N_c$.  
Without FI terms, this is a consequence of quantum effects.  It also
follows purely classically if we turn on a large FI term for all of
the $Q$ or all of the $\tQ$.

This leaves us with $\rank G \le N_f-N_c$, which combined with a
large FI term force either $Q_i$ or $\tQ_j$ to take maximal rank $N_c$.
The relations $G Q = \tQ G = 0$ generally do not force $G=0$.

We need to postulate a new superpotential which leads to the
combination of the relations $F=0$ and \rfe{newrel}. 
This can be done by postulating the sum superpotential
$$
W_{new} = W|_{Q_i\tQ_j = M_{ij}} + \tr M q \tq
$$
as in \cite{BP,CFIKV,FHHU}, which leads to $F=\frac{\p W}{\p\phi_\alpha} = 0$
and
\begin{equation}\label{eq:newrelone}
\frac{\p W}{\p M_{ij}} \equiv G^{ij} = q_j \tq_i ;
\end{equation}
\begin{equation}\label{eq:newreltwo}
M q = 0 ;
\end{equation}
\begin{equation}\label{eq:newrelthree}
\tq M = 0 .
\end{equation}

We now need to postulate a general map.  Suppose the FI terms are
positive and $Q$ is maximal rank (otherwise we reverse all the arrows).
We again use  \rfe{tildeBseq}\ to determine $\tq$ from $Q$, which
will have maximal rank, and \rfe{Mdef}\ to determine $M$ from $\tQ$.
As before, this clearly satisfies \rfe{newrelthree}.

We then use \rfe{newrelone}\ to determine $q$ from $\tq$ and $G$.  
This is an overdetermined system of equations, but given that $G$
has rank at most $N_f-N_c$ it will have a unique solution up to gauge
invariance.
The relation \rfe{newreltwo}\ then follows from $\tQ\cdot G = 0$.

Conversely, given $(G,q,\tq,M)$ we determine $Q$ from $\tq$ and $\tQ$
from $M$ as before.  $G \cdot Q=0$ then follows from \rfe{newrelone},
while $\tQ \cdot G = 0$ follows from \rfe{newreltwo}\ and \rfe{newrelone}.
We have not used $q$, but because of \rfe{newrelone}\ this is redundant
information if we know $G$.

Finally, since the map was one-to-one before putting the additional
constraints, then given that the images in both directions satisfy the
constraints, it must be one-to-one between the spaces satisfying the
constraints.

This shows that the new superpotential reproduces the original
moduli space.  In some cases one can then integrate out fields
to get a simpler theory with the same moduli space.

A special case of what we just discussed is the verification that
the duality is an involution in the usual way, by taking $G^{ij}$
constant.  This gives the electric theory with a mass term, and we
conclude that its magnetic dual is
$$
W_{new} = \tr G^{ij} M_{ij} + M q \tq
$$
We can then dualize again, and eliminate the additional field, to
arrive back at the original electric theory.  The additional ingredient
added by the above is the check that the map on configurations is the
identity map.

We finally turn to the case in which the superpotential is not linear
in $M$, and thus the matrix $G^{ij}$ depends on $M$.  Actually, for
the problem of finding classical supersymmetric vacua, this does not
make any difference.  What we have already proven is that we have a
one-to-one equivalence between configurations $(G,Q,\tQ)$ and
configurations $(G,M,q,\tq)$ for any $G$, in which $M=Q\tQ$.  The
effect of letting $G$ depend on $M$ is that supersymmetric vacua will
be a subset of pairs $(G,M)$ defined by this relationship.  
This subset is the same on both sides in the obvious way, so the same
map will be one-to-one in this case as well.

\subsection{Adjoint matter}

The previous arguments assumed that the nodes $F_i$ were all
distinct from $E$, in other words that there was no adjoint matter
associated to the node $E$, or $\Ext(E,E)=0$.

Suppose we have adjoint matter $\Ext(E,E)$ with no superpotential,
call it $\chi$.  A physical way to think about this is that the brane
$E$ now comes in a family with continuous parameters.  Call these
$E_\chi$ where $\chi$ can now be thought of as an eigenvalue of the
field $\chi$.  

If there is no superpotential, there will be bound
states between the branes $F_i$ and an arbitrary number of $E_\chi$
with differing $\chi$.  For example, we have the holomorphic configurations
\begin{equation}\label{eq:vevchi}
Q = (1\ 0\dots); \qquad
\chi \sim \begin{pmatrix} 0& 0 & 0& \dots\\
1&0&0&\dots\\
0&1&0&\dots\\
0&0&1&\dots\\
\vdots& \vdots& \vdots & \ddots
\end{pmatrix} 
\end{equation}
for arbitrary rank matrices.  These can also satisfy the D term constraints,
\begin{equation}
D_2 = Q^\dagger Q + [ \chi^\dagger,\chi] >0\label{eq:Dtermadj}
\end{equation}

So, there is no maximal bound state with which we can apply our
arguments.  At best, the style of argument we gave will lead to a
theory with infinite gauge groups.  Mathematically, this
is not a quiver of finite representation type, meaning (by definition)
that there are an infinite number of simple bound states.

This problem can be avoided by starting with a superpotential which
only allows a finite number of $E_\chi$, or a finite number of
bound states.  A case discussed in the literature is to
add the superpotential 
$$
W = \tr P(\chi)
$$
for some degree $k+1$ polynomial $P$  \cite{kutasov, KSS}.  Now there are $k$
supersymmetric configurations of the brane $E$, labelled by the roots
$\chi_i$ of $P'(\chi)=0$.

Let us now restrict attention to SQCD with this additional field; we
again dualize the $B_2$ node (so $E=B_2$).  Let $B_2^i$ with $1\le
i\le k$ be the $k$ different configurations of $B_2$.  If the roots
$\chi_i$ are all distinct, we can handle this case by successively
dualizing each of these branes along the same lines as before.

The new possiblility is that some of the roots $\chi_i$ coincide.  Let
us take $P=\chi^{k+1}$ for definiteness.  Now there are a series of
bound states $E_t$ with charges $(0\ t\ 1)$, with field configuration
\rfe{vevchi}.  Among these, one expects $E_k$ to be a maximal bound
state, which can play the role of $B_5$ above.

Changing basis from $(B_1,B_2,B_3)$ to $(B_1,\bar B_2,B_5)$ now leads
to $N'_{\bar 2}= k N_3 - N_2$, the same as the rank of the dual theory
found in \cite{kutasov,KS,KSS}.

We now need to derive the morphisms in the new basis.  We should also
verify that $B_5=E_k$ is maximal.  One could do this by hand as we did
before, but at this point some mathematical technology will be
welcome, so we postpone this to the next section.

\section{Dirichlet brane realization}\label{sec:Dbranecat}

The previous arguments can be embedded into the discussion of
D-branes in the weak string coupling limit, which provides
a complete definition including branes and antibranes.
Indeed, this is how they were found.  
We summarize the most relevant aspects of this theory from
\cite{DougDC} (other discussions of D-brane categories can be found in
\cite{AL,Diaconescu,Lazaroiu1}).

We consider type \IIb\ D-branes embedded in a Calabi-Yau threefold
$\CM$ and extending in the $3+1$ Minkowski dimensions, such that their
world-sheet theory has at most $\CN=1$ supersymmetry.  We say at most
because a generic collection of branes will break all supersymmetry.
This should be thought of as a spontaneous breaking, and the
(low energy) world-volume Lagrangian satisfies the constraints of $\CN=1$
supersymmetry.  

In translating from geometry to the $\CN=1$ language, the holomorphic
structure of the CY and brane embedding becomes holomorphic data
(spectrum and superpotential), while K\"ahler data is related to gauge
couplings, D-flatness conditions and FI terms.  The discussion is best
phrased in this two-step way, as we did in the previous section.  In
particular, most of the subtleties involving branes and antibranes
only appear in this second step.

In world-sheet terms, the holomorphic structure of the theory is all
visible in the topologically twisted open string theory.  D-branes are
boundary conditions in this theory.  The open string states are
precisely the massless Ramond states of the physical theory, while the
physical superpotential is the generating function for their
correlation functions.  The explicit definitions are simplest in the
B model in the large volume limit, in which boundary conditions are
holomorphic bundles, massless Ramond states are Dolbeault cohomology,
and the superpotential is the holomorphic Chern-Simons action.  Other
starting points such as $(2,2)$ CFT or orbifold CFT can be used as well.

Independent of the starting point, the structure of open string theory
allows a direct generalization to use complexes as defined in
\rfe{defcomplex}\ as boundary conditions, by including the information
of the $d$ operator in the BRST charge.  One can then argue that
all configurations related by brane-antibrane annihilation as in
\rfe{vacuum}\ are quasi-isomorphic, and that passing from the original
configurations to their derived category incorporates all possible
brane-antibrane bound states in a systematic way.

The choice of which of these objects are branes and which are
antibranes then depends on K\"ahler moduli.  The basic results
relevant for us are the following.  First, the holomorphic structure
of an object is independent of K\"ahler moduli, and this will guarantee
that we can take a specific configuration from one basis to another,
which was the underlying basis of our argument.

Next, let us compare a brane $B$ extending in the $3+1$ Minkowski
dimensions, with a brane in \IIa\ theory wrapping the same cycle (and
entirely the same in the CY), but which is a BPS particle in $3+1$
dimensions.  This particle will have a central charge $Z(B)$ depending
on K\"ahler moduli, and computable using mirror symmetry.  A brane and
its antibrane have $Z(B)=-Z(\bar B)$.  More generally problems in
which all $Z(B_i)$ are roughly aligned in the complex plane can be
treated using supersymmetric gauge theory of ``branes,'' while more
general combinations involve brane-antibrane subtleties.

The same central charge enters into many physical quantities in
the \IIb\ world-volume theory.  Let us first consider a
collection of branes whose central charges are all close to real
numbers (possibly after some overall phase rotation), we have
$$
Z(B) = \frac{1}{g_B^2} + i \zeta_B
$$
where $g_B$ is the Yang-Mills coupling for its $U(N)$ gauge group, and
$\zeta_B$ is its FI term (and we work with $\alpha'=1$).
This result has numerous consequences.  For example, the question of
whether the boson partner to a massless fermion is tachyonic or massive,
and thus whether the two branes form a bound state,
is determined by the relative phase of the central charges.

The condition that central charges are almost real is not necessary; 
the more general result is
\begin{equation}\label{eq:zcharge}
\frac{1}{g_B^2} = |Z(B)|; \qquad
\zeta_B = \grade(B) = \frac{1}{\pi} \arg Z(B).
\end{equation}
We will use the notation $\grade(B)$ or ``grade'' below as it is more
precise (the relation to $\zeta_B$ is only precise in field theory),
and explain how this enters our considerations.

Now, it is clear from the brane discussions \cite{EGK,EGKRS} that Seiberg
duality provides a field theory description of the result of going
through ``infinite gauge coupling.''  This corresponds to going through
zero volume or $Z(B)=0$ in the present discussion, which is singular
in CFT and of course takes us out of weak string coupling.  

Of course, because the central charge is complex, we can avoid this
region, going around a point $Z(B)=0$ on a path with $|Z(B)|>>1$.
In field theory, this amounts to turning on large FI terms, as in our
previous discussion.  The result takes (say) positive real $Z(B)$ to
negative real $Z(B)$.  To turn this into a problem involving only
branes and thus describable by supersymmetric gauge theory, we instead
use $\bar B$, which by the discussion we just made now counts as a
{\it brane}: $Z(\bar B)>0$.  We then need to do one of the changes of
basis discussed above to get a new basis whose objects have minimal
$Z(B'_i)$, and in terms of which the configuration is again a bound
state of positively many constituents.

Note that in this discussion, there are two paths by which one can
continue around $Z(B)=0$, taking $\Im Z(B)>0$ or taking $\Im Z(B)<0$.
These are distinct physical operations and correspond to the two
changes of basis \rfe{tildeBseq}\ and \rfe{tildeAseq}.  Of course they
lead to the same point in K\"ahler moduli space and thus the same dual
theory.  However they differ by a full loop around $Z(B)=0$, in other
words by a {\it monodromy} in the K\"ahler moduli space.  Such a
monodromy relates different presentations of the same physical theory
in terms of different bases of branes, and thus the set of physical
configurations of the two theories must be precisely the same.

This should be contrasted with the Seiberg duality itself which does
vary the K\"ahler moduli and thus need not preserve the entire
spectrum of classical solutions of F-flatness and D-flatness.  We
concentrated on the subset of configurations which are preserved, but
in general there are others which only exist on one side of the
duality.  The situation becomes more symmetric after considering
quantum effects, as we discuss below.

In the language above, the Seiberg dualities arise on performing
``partial'' monodromies.  As we discuss below (and as pointed out in
\cite{CFIKV}), in the CY examples, these can be identified with
mutations on the exceptional collection of bundles corresponding to
the fractional branes.

\subsection{D-brane categories}

Let us consider two D-branes $B_1$ and $B_2$ in a smooth Calabi-Yau
manifold $X$ which are BPS in the large volume limit.  The simplest
case to start with is branes $B_1$ and $B_2$ wrapping $X$, with a
holomorphic vector bundle on its world volume.  The spectrum of
massless fermionic strings stretching between branes $B_1$ and $B_2$
is then given by elements of Dolbeault cohomology,
$H^{0,q}(X,B_1^\dual\otimes B_2)$.  These are just holomorphic
$q$-forms, which can be multiplied (by wedge product).

We can next add to this branes wrapping lower dimensional cycles
carrying bundles.  However the set of these does not form an abelian
category, because kernels and cokernels of maps between bundles are
not necessarily bundles.  Rather, a more general and algebraic
description is that the branes $B_i$ correspond to coherent sheaves on
$X$.  This is a standard mathematical construction (e.g. see
\cite{GH,GM}) which includes the original bundles
(as locally free sheaves of sections of bundles),
bundles with singularities, and
objects supported on lower dimensional holomorphic cycles.
The spectrum of massless fermionic strings stretching between branes
$B_1$ and $B_2$ is then given by elements of $Ext^q(B_1,B_2)$.

The coherent sheaves form an abelian category and in this sense are a
``complete'' set of objects which one expects to appear as branes at
large volume.  Superficially, this looks very different from the
abelian categories of quiver representations discussed earlier.  On a
deeper level, they are very similar.  Indeed, one can find quiver
categories which ``represent'' them in the sense that large subcategories 
of the two are literally the same, the original example being
\cite{Beilinson}.  We will come back to this point later as it is quite
relevant.

The further discussion relies on properties of the world-sheet $\CN=2$
algebra.  Each open string has a $U(1)$ charge $Q$ (in the
topologically twisted string, this is the ghost charge).  For branes
which are bundles on $X$, this is just $q$ in $H^{0,q}$.  the
physical string, this sector is obtained by acting with $q$ fermions
$\psi^i_0$ on the ground state.  On its NS partner state, these are
$\psi^i_{-1/2}$, so the mass of the NS state is $m_{NS}^2=(q-1)/2$.

We also have to remember that in the physical theory we need to impose
the GSO projection, and that there are also fermionic oscillators
transverse to the Calabi-Yau manifold.  At this point we can
distinguish branes and antibranes.  For a pair of branes, elements of
$Ext^1(B,B)$ give rise to massless matter.  Geometrically, these
deform the connection of the associated holomorphic bundle, and
explore the moduli space of coherent sheaves.  Elements of $Ext^0=
\Gamma(Hom(B,B))$ are global holomorphic sections of the sheaf.  To
get a physical vertex operator, the GSO projection forces us to act
with an additional oscillator transverse to the Calabi-Yau, so these
give rise to vector particles in $R^{3,1}$, with $m^2=q/2$.

For a brane and antibrane, the GSO projection is reversed.  Thus an
$\Ext^0(B_1,B_2)$ becomes the tachyon between a brane and anti-branes
(this state is projected out by the GSO projection for a brane), now
with $m^2=(q-1)/2=-1/2$ for $q=0$.

The category of sheaves also has the Serre duality functor, 
which on a Calabi-Yau $d$-fold takes the form
\begin{equation}\label{eq:serredual}
\Ext^q(A,B) = \Ext^{d-q}(B,A)^*.
\end{equation}
In each open string sector, the GSO projection will eliminate half 
of the states, either the even or odd $q$.  Since for us $d=3$ is odd,
we can implement this by considering both $\Ext(A,B)$ and $\Ext(B,A)$,
but restricting attention to the states with $q<d/2$.

We represent the resulting chiral matter spectrum 
in a quiver diagram as follows.  Each brane is represented by a node.
Between two branes $A$ and $B$, we have an arrow 
$A$ to $B$ for each basis vector of $Ext^{odd}(A,B)$, 
and an arrow from $B$ to $A$ for each basis vector of $Ext^{even}(A,B)$.

All this assumed that $U(1)$ charges are integral.  However, a general
combination of branes, breaking supersymmetry, will have open strings
with non-integral $q$.  This is true even in the large volume limit,
as can be seen by considering the system of a D$p$ and D$p-2$-brane.
This is because $q$ and $m^2$
receive a correction related to the NS ground state energy (for
example, two branes with real relative dimension $n$, i.e. $n$ ``DN''
boundary conditions, have a vacuum with $m^2=n/4-1/2$).  This effect
can be summarized in terms of the gradings of the branes, $\grade_i =
\frac 1\pi \arg(Z(D_i))$: one then has $q=q_0+\grade_2-\grade_1$, where
$q_0$ is the grading of the $\Ext^{q_0}$ in the standard definitions.

We will mostly be concerned with field theory limits, in which $q$ is
always near an integer, but let us discuss the general case.
The formula $m_{NS}^2=(Q-1)/2$ still applies;
for example, the D$p$-D$p-2$ tachyon has $m^2=-1/4$, and $q=1/2$.  
In general, the $U(1)$ charges at a given
point in K\"ahler moduli space can be obtained by ``flow of gradings.''
Between a point $x$ with gradings $\grade_x$ and another point $y$
with gradings $\grade_y$, writing $\Delta\grade(B)=\grade_y(B)-\grade_x(B)$,
the charges are related as
$$
\Ext^q(B_1,B_2) \rightarrow
 \Ext^{q+\Delta\grade(B_2)-\Delta\grade(B_1)}(B_1,B_2) .
$$
The GSO projection does not vary along such a flow.  A more convenient
notation for the same thing is
$$
\Hom(B_1,B_2[q]) \rightarrow
 \Hom(B_1[\Delta\grade(B_1)],B_2[\Delta\grade(B_2)]) .
$$

This leads to the corresponding variation
$$
m^2 \rightarrow m^2+\Delta\grade(B_2)-\Delta\grade(B_1)
$$
(just as for varying FI terms) and this can in general lead to
$m^2<0$ if the phases $\grade(B_i)$ vary appropriately. 
Therefore, via motion in moduli space, a massive field can become
massless or tachyonic, which 
can force one collection of D-branes to condense into a bound state.

We will use this to infer the result of a flow in moduli space along
which a brane ``turns into an antibrane,'' more precisely along which
$Z(E)$ goes from positive real to negative real.  As we discussed, this
can happen in two ways, which take $\grade(E)\rightarrow\grade(E)+1$
or $\grade(E)\rightarrow\grade(E)-1$.  Let us consider the first; then
matter $\Ext^1(E,B_i)$ turns into ``partial gauge invariances''
$\Hom(E,B_i)$, which are somewhat problematic from the point of view
of gauge theory.  The antibrane $\bar E$ has spectrum obtained by
reversing the GSO projection; the same maps are interpreted as
tachyonic matter $\Hom(\bar E,B_i)$, which in some sense
lead to the formation of bound states which make up the new basis.

In the new basis, the shift takes $\Hom(B'_i,E)$ to
$\Ext^1(B'_i,\bar E)$, and shows that this dual quark is a massless
chiral multiplet.  Similarly the original $\Ext^1(B,E)$ for a brane
$B$ not transformed under the duality becomes $\Ext^2(B,\bar E)$, which
is massive; its Serre dual $\Ext^1(E,B)$ would become a massive vector,
but its antibrane version $\Ext^1(\bar E,B)$ is a chiral multiplet,
the other dual quark.

\subsection{D-branes and quiver categories}

Given two branes $B_1$ and $B_2$ defined as sheaves, quiver representations,
or whatever, the $\Ext^1(B_2,B_1)$ group classifies all of the 
branes which can be defined via short exact sequences
\begin{equation}
0\to B_2\mapr^f B \mapr^g B_1 \to 0\label{eq:bound}
\end{equation}
We think of $B$ as a deformation of $B_2\oplus B_1$, given by 
condensing a chosen field in $\Ext^1(B_1,B_2)$.  The arrows
here are $f\in\Hom(B_2,B)$ and $g\in \Hom(B,B_1)$.  They reflect the
fact that $B_1$ and $B_2$ are constituents of $B$, and that this can
be seen in the appearance of ``partial gauge invariances.''

In section \ref{sec:chabas} we implicitly used exact sequences as
above for our computations. Indeed, the results presented there can be
rewritten in this language, where the procedure becomes more
natural.  The main computational tool in this description is the
long exact sequence in cohomology
\begin{eqnarray}
\dots 0\to& \Ext^0(M,B_2) \to \Ext^0(M,B) \to \Ext^0(M,B_1)&\\
\to & \Ext^1(M,B_2) \to \Ext^1(M,B) \to \Ext^1(M,B_1) &\dots  .
\end{eqnarray}

This may be familiar to the reader from algebraic topology, and
as an abstract statement it works the same way here \cite{Osborne}.  There is a
similar sequence for $\Ext(B,M)$ obtained by reversing arrows.
\begin{eqnarray}
\dots 0\to& \Ext^0(B_1,M) \to \Ext^0(B,M) \to \Ext^0(B_2,M)&\\
\to & \Ext^1(B_1,M) \to \Ext^1(B,M) \to \Ext^1(B_2,M) &\dots  .
\end{eqnarray}

We are going to use this shortly, but first let us remark on the
underlying definitions.  The overall context of this section is B-type
brane realizations in CY, for which the underlying definition of the
abelian category is sheaf cohomology.  The category of coherent
sheaves is an abelian category with Serre duality, and the
``decoupling statement'' of \cite{BDLR} (or general considerations of
topological open string theory, if one likes) tells us that this is
the abelian category containing the bundles we started with.

One can make the following arguments precise by choosing a specific
geometry and wrapping D-branes to realize the following quiver
theories on their world-volumes; the $\Ext$ groups we discuss are then
those of sheaf cohomology.  To be a bit more precise, suppose we are
interested in a quiver theory with superpotential, call it $T$.  Let
$\Coh M$ be the category of coherent sheaves on a Calabi-Yau $M$.  We
choose some $M$ such that there is some full subcategory $C(T)\subset
\Coh M$ in which the branes of our discussion are contained, meaning
that the relevant physical computations (e.g.  spectrum and
superpotential) reproduce $T$.  Then, we are working in $C(T)$.  Of
course there might be many $C(T)$'s which could play this role.

We are not going to make this step explicit, because our philosophy is
somewhat different.  Rather, a quiver theory $T$ {\it by itself} gives
rise to an abelian category, call this $\Cat T$.  For the theories
that arise from geometry, this will be the same as the geometric
category $C(T)$, no matter what geometry we derive it from.  A
particular quiver theory might {\it a priori} not even have a
geometric realization.  It is this category $\Cat T$ which we have in mind in
the following discussion.  If we believe this claim, then the
dualities we are discussing again do not follow from string theory or
even geometry, but are simply properties of supersymmetric field
theory.

We will explain this point in section 5.  If one is familiar with the
geometric approach, one can keep a specific $C(T)$ in mind in reading
the following arguments.  If one is familiar with quivers, one may be
bothered by the fact that these categories do not in general have
Serre duality.  However, we will show in section 5 that if the
relations follow from a superpotential, then under certain assumptions
the category will have Serre duality.

\subsection{Seiberg duality via long exact sequences}\label{sec:seidualrev}

We now return to the theory $(N_1,N_2,N_3)$ of section $2$.  
We recall that $\dim(\Ext^1(B_2,B_3))= \dim(\Ext^1(B_1,B_2))= 1$, and
$\dim(\Hom(B_i,B_j))= \dim(\Ext^0(B_i,B_j)) = \delta_{ij}$.

The bound state $B_5$ between branes $B_2, B_3$ is 
defined by a short exact sequence $0\to B_3 \to B_5\to B_2\to 0$. 
Since there is only one matter field, $B_5$ has no moduli;
in other words there is a unique non-trivial extension.

Now, let us calculate the spectrum between brane $B_1$ and brane $B_5$.
Chasing the long exact sequence in homology we find
\begin{equation}
\xymatrix{0\ar[r] & 0 \ar[r] & \Hom(B_1,B_5) \ar[r] & 0 \mapr \\
& 0 \ar[r] & \Ext^1(B_1,B_5) \ar[r]& \Ext^1(B_1,B_2)\mapr\\
& 0 \ar[r] & \Ext^2(B_1,B_5) \ar[r]& 0 \mapr\\
& 0 \ar[r] & \Ext^3(B_1,B_5) \ar[r]& 0 }
\end{equation}
So we find that $\dim(Ext^1(B_1,B_5)) = 1$, and that it is derived
from $\tilde Q\in\Ext(B_1,B_2)$. 
This is of course the meson $M$ in the dual theory.

To get $\Ext^p(B_5,B_2)$, we can use the long exact sequence
\begin{equation}
\xymatrix{0\to Hom(B_2,B_2) \ar[r] & \Hom(B_5,B_2) \ar[r] & 0\\
0 \ar[r] & \Ext^1(B_5,B_2) \ar[r]& 0 \mapr\\
 0 \ar[r] & \Ext^2(B_5,B_2) \ar[r]& \Ext^2(B_3,B_2) \mapr\\
 \Ext^3(B_2,B_2) \ar[r] & \Ext^3(B_5,B_2) \ar[r]& 0 }
\end{equation}
We want to see that the nature of the extension is such that there is a 
connecting homomorphism between lines $3$ and $4$, so we find that in the end
 the
only non-trivial ext group is $Hom(B_5,B_2) \simeq Hom(B_2,B_2)$.

This is most easily seen with the dual exact sequence
\begin{equation}
\xymatrix{0\ar[r] & 0 \ar[r] & 0 \ar[r] & \Hom(B_2,B_2) \mapr \\
& \Ext^1(B_2,B_3) \ar[r] & \Ext^1(B_2,B_5) \ar[r]&0\mapr\\
& 0 \ar[r] & \Ext^2(B_2,B_5) \ar[r]& 0 \mapr\\
& 0 \ar[r] & \Ext^3(B_2,B_5) \ar[r]& \Ext^3(B_2,B_2) } .
\end{equation}
Actually both of these sequences contain the same information,
all the groups are related by Serre duality.
However, in this second exact sequence we know that we are using
a non-trivial (canonical) extension which identifies 
$Hom(B_2,B_2)\sim Ext^1(B_2,B_3)$. This is in accordance with the field 
theory: giving a vev to a field reduces the number of massless degrees of
freedom.

In the bound state $B_5$, the connecting map pairs $\Ext^1(B_2,B_3)$
with $\Hom(B_2,B_2)$, leaving $\Ext^1(B_2,B_5)=0$.  This is really the
same argument we gave in section 2, that the extra matter is Higgsed in
this case.  If one had considered the trivial extension $B_2\oplus B_3$,
the connecting map would have been zero, and both gauge and matter
fields would remain massless.

Dualizing everything, one finds that $\dim\Hom(B_5,B_2)=1$,
as before.  In fact $\Hom(B_5,B_2) \cong \Hom(B_2,B_2)$.

Now we use this to obtain the morphisms to the antibrane of $B_2$,
defined by shifting the ghost charge by $\pm 1$.  By the general
principle that gradings are non-negative, we must take $\bar B_2=B_2[-1]$,
i.e. shift its ghost charge by $-1$.  Thus this morphism is also an
$\Ext^1(B_5,\bar B_2)$.
This should be drawn as an arrow begining at $B_5$ and ending at $B_2$, so 
we see that we reverse the arrows in the quiver. 
This shift also relates $\Ext^1(B_1,B_2)$ to $\Ext^2(B_1,\bar B_2)$ which
is dual to $\Ext^1(\bar B_2,B_1)$.  

In terms of variation of K\"ahler moduli, this shift is induced by
varying the central charge $Z(B_2) \sim e^{i\pi\grade}$ from $\grade=1$
to $\grade=0$ above $Z=0$.  The path below $Z=0$ would produce 
$\bar B_2=B_2[1]$ and we would need to use $B_4$ instead of $B_5$
to get a sensible basis.

For completeness, we should check that $B_5$ is simple, {\it i.e.}
$\Ext^1(B_5,B_5)=0$.  This can be done by a succession of exact
sequences, but follows more simply from the original gauge theory
definitions.

\begin{figure}[ht]
\begin{center}
\leavevmode
\epsfxsize=7cm
\epsfbox{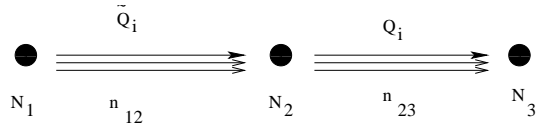}
\end{center}
\caption{Quiver with more arrows}
\label{fig:quiver4.eps}
\end{figure}

Let us now consider a slightly more general case, with three nodes 
$B_1,B_2,B_3$, but now with $n_{12}$ arrows from node $B_1$ to $B_2$ and
$n_{23}$ arrows departing from node $B_2$ to node $B_3$, as shown in the 
figure \ref{fig:quiver4.eps}.

Again, the basic duality will be obtained by taking $Z(B_2)\mapr
-Z(B_2)$.  We then want to go to a new basis including $\bar B_2$ and
a supersymmetric bound state $B_5$ with $(0,m,1)$ with $m$ as large as
possible.  This could be justified by asking for branes with minimal
central charge, or equivalently asking that all central charges can be
realized as positive combinations of those from the basis.

Clearly $B_5=(0\ n_{23}\ 1)$ as for larger $m$ there are not enough
matter fields to break $U(m)$ gauge symmetry.  The same constraint
arises by asking to solve the D-flatness conditions with large FI
terms.

There is a unique bound state of this form, given
by the tautological exact sequence
\begin{equation}
0 \to  Ext^1(B_3,B_2)^* \otimes B_2 \to B_5 \to B_3 \to 0 .
\end{equation}
In components, we set the fields $Q^i_\alpha=\delta^i_\alpha$, where
$1\le i\le n_{23}$ is the flavor index and $\alpha$ is the gauge index.

Chasing diagrams, we find that $\dim(Ext^1(B_5,B_1))= n_{12}n_{23}$, and that 
all other arrows are reversed with respect to the orginal diagram.

The same ideas can be pushed to handle the general theory with
bifundamentals and no superpotential, reproducing the results for
dualizing a node.  The main point where input about the specific
branes enters is in the connecting homomorphisms.  As we saw, the maps
$\Hom\mapr\Ext^1$ and the dual $\Ext^2\mapr \Ext^3$ 
correspond to the Higgs mechanism as giving mass to
the `extra' quark field.  

\subsection{Connecting maps from the superpotential}

The connecting maps $\Ext^1\mapr\Ext^2$ correspond to lifting matter
by the superpotential.  As a simple illustration, we check that
applying the duality twice returns us to the same theory.

Actually, we had two derivations of the duality, distinguished by
whether $B_2$ bound with $B_3$ or $B_1$.  As we explained, these were
associated to the shifts $B_2\mapr B_2[-1]$ and $B_2\mapr B_2[1]$,
respectively.  The more interesting case is to shift again in the same
direction, say to $B_2[-2]$, which turns the half monodromy into a
full monodromy.  This is done by changing to a basis with a brane
$B_7$ defined by the exact sequence $ B_1\mapr B_7 \mapr B_2[-1]$,
while we keep brane $B_5$.  At the level of brane charge we 
have that $[B_2]\to [B_2]$, $[B_3]\to [B_3]+[B_2]$, 
and $[B_1]\to[B_1]-[B_2]$.

The nontrivial part of this test is that the dual mesons must be
lifted in the double dual theory. Also, the dual theory has a 
superpotential which affects the massless spectrum of the theory
 when we give vevs.
This can be seen by considering the arrows between brane 
$B_5$ and brane $B_7$.
This gives us 
\begin{equation}
\xymatrix{0\ar[r] & 0 \ar[r] & \Hom(B_5,B_7) \ar[r] & 0 \mapr \\
& 0  \ar[r] & \Ext^1(B_5,B_7) \ar[r]& \Ext^1(B_5,B_2[-1]) \mapr\\
& \Ext^2(B_5,B_1) \ar[r] & \Ext^2(B_5,B_7) \ar[r]& 0 \mapr\\
& 0 \ar[r] & \Ext^3(B_5,B_7) \ar[r]& 0 }\label{eq:monodromy}
\end{equation}
If things worked in parallel to
the first duality, the dual quarks $\Ext^1(B_5,B_2[-1])$ would give
rise to doubly dual mesons $\Ext^1(B_5,B_7)$.  However, the dual
theory has nonzero $\Ext^2(B_5,B_1)$, the Serre dual to its mesons,
and $\Ext^1(B_5,B_2[-1])$ instead pairs through the connecting map to these.
Thus there are no dual dual mesons.

In fact, we recover the original quiver diagram, but with basis a 
different set of branes. In mathematical terms, dualizing twice is
an autoequivalence of the derived category.

\subsection{Generating the superpotential}

The previous argument was nice, but it would be better to just derive
the dual superpotential.  It turns out that this can be done directly if
we have a realization of our quiver category which has Serre duality.

If we have realized our branes as B-type D-branes in a CY, we can base
this on the standard result that the superpotential is the holomorphic
Chern-Simons action \cite{Wittentop},
\begin{equation}\label{eq:holocs}
W = \int \Omega \wedge \left(\bar A\bp \bar A + \frac{2}{3}\bar A^3\right).
\end{equation}
For holomorphic forms $\bar A$, the derivative term can be dropped.

In the notations we are using now, we regard $A$ as a sum of elements
in $\Ext^1(A,B)$ for all $A$ and $B$.  Since these are morphisms in
a category, we know how to multiply them.  We then write
$$
W = \tr A^3
$$
where $\tr \Ext^3(A,A) =1$ for all $A$ and is zero on anything else.
This formula should be interpreted in the following sense:
$\Ext^3(A,A) = \Hom(A,A)^*$. Now, in $\Hom(A,A)$ we have the identity 
homomorphism $1_A$, so we should interpret $\tr(a)$ for
$a\in \Ext^3(A,A)$ as given by the canonical pairing 
$\Hom(A,A)\times \Hom(A,A)^* \to \BC$, 
$<1_A, a>$, which is a complex number.

We then need to adapt this to a system of branes and antibranes.
Actually, if we identify an antibrane as a brane with shifted grading,
such that the morphisms are the same with shifted grading, then
the same formula applies, since the superpotential is independent of
these shifts (they only affect K\"ahler moduli).  The only
generalization we need is to allow $\bar A$ to be an arbitrary sum of
$p$-forms.  (See \cite{Vafa,Diaconescu} for \rfe{holocs} written 
out explicitly this
way, but it is just standard mathematical formalism.)  
So, if we had a term involving three $\Ext^1$'s like
$$\int \bar A_{1,2} \wedge A_{2,3} \wedge A_{3,1},$$
shifting the grade of $B_2$ produces a term 
$\Ext^2 \times \Hom \times \Ext^1$, etc.\footnote{
There is a slightly confusing point here which is that the
coefficients stay the same, which means that we can violate the
naive sign conventions.  For example, if this product started out
antisymmetric (as one would expect for one-forms), it stays antisymmetric
when shifted to even forms.  This is not contradicting anything as the
algebra is not commutative or supercommutative in general, but should
be kept in mind.}

\begin{figure}[ht]
\begin{center}
\leavevmode
\epsfxsize=7cm
\epsfbox{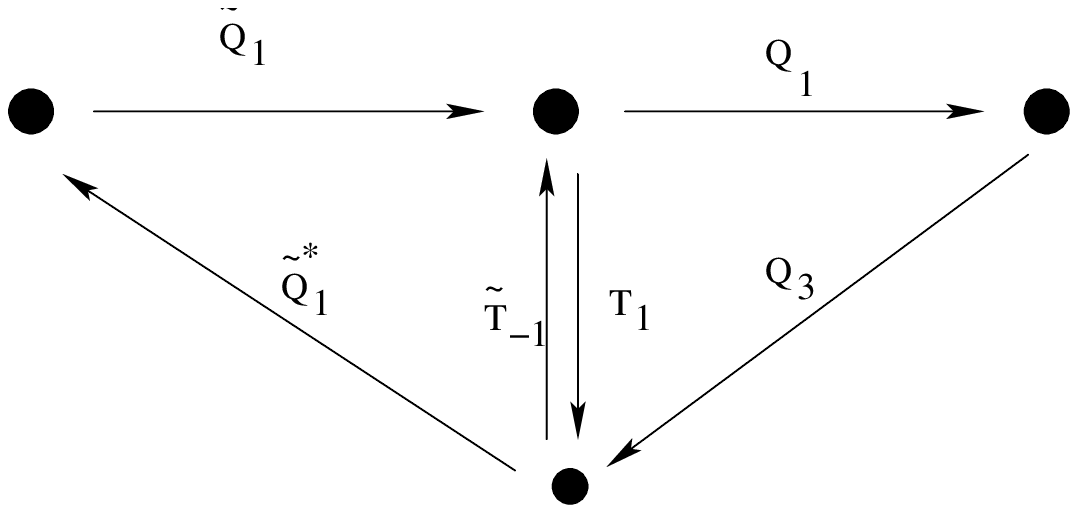}
\end{center}
\caption{Quiver with antibranes. The ghost charges of 
the arrows are explicitly written as a subindex. } 
\label{fig:quiver3.eps}
\end{figure}

The most direct application of this is to derive cubic terms involving
a brane-antibrane tachyon.  We work with our SQCD quiver but now
include the antibrane to $B_2$, defined $\bar B_2=B_2[-1]$, as in
figure \ref{fig:quiver3.eps}.

The quiver contains two tachyons, which are shifts of $Ext^{0,3}(B_2,B_2)$.
One is $T \in \Ext^1(B_2,B_2[-1])$, and the other is 
$\tilde T \in \Ext^{-1}(B_2[-1],B_2)$.

We also include the fields $Q_{0} \in \Ext^3(B_3,B_2[-1])$
(the shift of the Serre dual of $Q$, $Q^\dual\in\Ext^2(B_3,B_2)$. It's dual is
$Q_3$ in the figure, keeping with the convention 
that we only have $Ext^{odd}$), and
$\tQ_{2} \in \Ext^1(B_2[-1],B_1)$, the shift of the Serre dual
${\tQ}^\dual \in \Ext^2(B_2,B_1)$.

By Serre duality, the original superpotential included the terms
$\tr Q \cdot 1 \cdot Q^\dual$ and
$\tr \tQ^\dual \cdot 1 \cdot \tQ$.
The shift turns these into
$$
W = \tr Q \tT Q_3 + \tQ_1^* T \tQ_1 .
$$
with the notation as in figure \ref{fig:quiver3.eps}.

Since the ``tachyon'' is just obtained by shifting the unit operator,
it enters in a canonical way.

Looking back at our derivation of the dual spectrum, we can
identify $\tq$ as a tachyon between $\bar B_2$ and the $B_2$ constituent
of $B_5$, and $M$ as a string from $B_1$ to the $B_2$ constituent of
$B_5$, leaving $q$ as its ``Serre dual'' and justifying the identification
of one of these cubic terms with our dual superpotential.

This argument can be made a little more directly by just computing
$$
\tr \Ext^1(B_1,B_5) \cdot \Hom(B_5,B_2) \cdot \Ext^2(B_2,B_1)
$$
as a product of linear transformations between the original quiver
representations.  Tracing through the definitions, one sees that
$$
\Ext^1(B_1,B_5) \cdot \Hom(B_5,B_2) \in \Ext^1(B_1,B_2) ,
$$
is nonzero (it is the original defining $\Ext$).  By definition this
has a nonzero trace with its Serre dual.  Shifting leads directly to
the superpotential in the dual theory.

To repeat, understanding Serre duality is the key to getting the
superpotential in these arguments.  We will argue in section 5 that
even this property of branes on CY is not necessarily geometric, but
is a more general property of quiver gauge theories.

\subsection{Dualizing theories with adjoints}

We are now prepared to consider the situation where we take the
original example we started with and we add an adjoint superfield for
node $B_2$. Here we expect new phenomena, based on the results
obtained by \cite{kutasov, KSS}, who found a nontrivial relation
between the original and dual superpotentials.

As discussed earlier, we need a superpotential, and we use
$W=\frac 1{k+1}\tr(\chi^{k+1})$, which forces $\chi^k = 0$.
We then described a candidate maximal bound state $B_5$, with
$[B_5]=k[B_2]+[B_3]$, as an explicit field configuration.
This configuration will be the bound state of $B_3$ 
with the maximum possible number
of branes of type $B_2$.

In the language of exact sequences, $B_5$ can be built by a sequence
of extensions, successively producing the bound states $E_t$ with
charges $(0\ t\ 1)$, starting with
\begin{equation}\label{eq:extstep1}
0\to B_2 \to E_1 \to B_3\to 0 .
\end{equation}
The exact sequences we used earlier now show that
$\dim\Hom(B_2,E_1)=1$ as before, but also
$\dim(\Ext^{1,2,3}(E_1,B_2))=1$.
This is just the statement that the brane $B_2$ inside $E_1$
has an adjoint which does 
not get lifted when we give a vev to $Q$. In the exact sequence this 
is read from $\Ext^{1,2}(E_1,B_2) \simeq \Ext^{1,2}(B_2,B_2)$

The best way to think about this calulation is by considering the system 
as describing the arrows between branes $(0,1,1)$ and $(0,1,0)$; so it 
arises from the total brane system $(0,2,1)$. When we turn on the 
vev for the quark field, 
it breaks the $U(2)\times U(1)$ symmetry to a 
$U(1)\times U(1)$ symmetry. We have an action of the $U(2)$ on a 
vector space of dimension 2, and let us label it's basis as
 $e_1, e_2$. We choose $e_1$ so that it 
spans the image of the quark field when considered 
as a map between the vector space 
associated to brane $B_3$, and the one associated to $B_2$.
A basis of $\Ext^{1,2}(B_5, B_2)$ counts massless fields transforming 
as bifundamentals between the gauge groups of the bound state brane $B_5$ 
and $B_2$. These are the off-diagonal matrix elements of the field $\chi$.
One can not turn on the element $\chi_{11}$ because the superpotential 
forbids it. However this field is not massive. The obstruction is at a 
higher order. We will choose now to set $\chi_{ii} =0$ for all 
diagonal elements (eventually all of these fields will become massive, 
but for the time being this follows from enforcing the equations of 
motion $\chi^k=0$).

We can now continue, with either $\Ext^1(E_1,B_2)$ or
$\Ext^1(B_2,E_1)$.  This amounts to the choice of which
matrix element of $\chi$ we turn on, $\chi_{1,2}$ or $\chi_{2,1}$.
The choice we take is determined by the structure of the D-terms, as these
two will contribute with opposite signs. In our conventions we will turn 
on $\chi_{2,1}$. The way extensions are made succesively is determined by
the first step \ref{eq:extstep1}, we need to keep extending with brane $B_2$
allways on te same side of the exact sequence.
In any case, 
we will have an extension (either to the left or right) 
as follows
\begin{equation}
0\to B_2 \to E_t \to E_{t-1}\to 0 .
\end{equation}

Since $\Ext^1(E_1,B_2)\neq 0$, we can find a non-trivial canonical extension
$0  E_2 \to E_1\to B_2\to  0$, which has charge $B_3 +2 B_2$. 
In general we can build successively $E_t$ via a sequence when
$\dim(\Ext(E_t,B_2))\neq 0$. As we have argued already, chasing the exact 
sequences will take care of both the Higgs mechanism, and of terms that 
appear in the superpotential. It can be proven that for general $E_t$ with 
$t<k$, that $\dim(Ext^{1,2,3}(E_t,B_2)) =1$. 
 The exact sequence will look like
\begin{equation}
\xymatrix{ 0 \ar[r]& \Hom( E_t,B_2)\ar[r] & \Hom(B_2,B_2)\to\\
Ext^1( E_{t-1}, B_2) \ar[r] &  Ext^1( E_t,B_2)\ar[r] & Ext^1(B_2,B_2)\to\\
Ext^2( E_{t-1}, B_2) \ar[r] &  Ext^2( E_t,B_2)\ar[r] & Ext^2(B_2,B_2)\to\\
Ext^3( E_{t-1}, B_2) \ar[r] &  Ext^3( E_t,B_2)\ar[r] & Ext^3(B_2,B_2)}
\end{equation}
If we write just the dimension of the vector spaces involved we get
\begin{equation}
\xymatrix{ 0 \ar[r]& \Hom( E_t,B_2)\ar[r] & 1\to\\
1 \ar[r] &  Ext^1( E_t,B_2)\ar[r] & 1 \to\\
1 \ar[r] &  Ext^2( E_t,B_2)\ar[r] & 1 \to\\
1 \ar[r] &  Ext^3( E_t,B_2)\ar[r] & 1}
\end{equation}
The canonical extension provides us with a connecting homomorphism 
between lines one and two, which can be traced to the higgs 
mechanism. This should also apply between lines three and four, because 
$Ext^3(E_t,B)$ comes from the vector fields associated to the 
$B_2$ branes inside $E_t$. However, we can only get a connecting 
homomorphism between lines two and three if they arise from the 
superpotential of the theory.

In terms of a basis, we need to consider for $E_t$ the 
theory $(0,t-1,1)+ (0,1,0)$. The extra field takes the basis
$e_1, e_2, \dots, e_{t-1}$ and adds one extra vector. 
Now we need to turn on an extension, which  corresponds to the
$\chi_{t,\alpha}$ components, as these are the fields
that transform as bifundamentals of the unbroken gauge group
for the  $E_t+ B_2$ brane configuration.
 However, for $\alpha<t-1$ these can be gauged away, so the only 
extension up to gauge invariance is given by the field $\chi_{t,t-1}$
which we can set equal to one. Notice how the extension is
providing the field configurations written in equation 
\rfe{vevchi}.

The sequence above tells us that for $E_t$ there is one degree of freedom
comming from $\chi$ which does not participate in the higgs mechanism.
We need to know when this particular degree of freedom needs to be
 integrated
out. Each time we perform a step, we add a column and a row 
to the matrix $\chi$.
The
quadratic spectrum of fluctuations receives a mass term
$\tr(\chi^{k+1}) \sim \tr(<\chi>^{k-1} \delta\chi\delta\chi)$ 
exactly when we reach the $k$-th step in the extension.
One can see that this happens
 because a further extension will give us a matrix $\chi$ 
such that $\chi^{k}\neq 0$, thus one would not be able to 
satisfy the F-term 
constraints and then the associated field is  massive.
At the $k$-th step
we find that $\Ext^{1,2}(E_k,B_2) = 0$, and it is not possible to do any
 further 
extensions; we have arrived at a situation where
the only non-trivial Ext
group is $\Ext^3(E_k,B_2)$.

 When we use this particular D-brane and shift to
obtain $\bar B_2$, we see that in the dual
 quiver the arrows 
between branes $B_5$ and $B_2$ are reversed with respect to those of
 the original 
quiver, between branes $B_2$ and $B_3$.
  The same is true for the arrows between $\bar B_2$ and $B_1$. 
Again, $B_1$ can
 not be extended 
with the choice of D-terms we made, because the quarks contribute with
 opposite sign to the D-term 
constraint, and it is not possible to get a possitive definite 
matrix. This statement is reflected in the exact sequences because we
can not extend $B_1$ to the left with copies of $B_2$, however we can 
extend it to the right.

Now, let us analyze the mesons. We want to use the exact sequence for $E_t$
 as follows

\begin{equation}
\xymatrix{ 0 \ar[r]& \Hom( E_t,B_1)\ar[r] & 0 \to\\
\Ext^1( E_{t-1}, B_1) \ar[r] &  \Ext^1( E_t,B_1)\ar[r] & \Ext^1(B_2,B_1)\to\\
\Ext^2( E_{t-1}, B_2) \ar[r] &  \Ext^2( E_t,B_2)\ar[r] & 0 \to\\
\Ext^3( E_{t-1}, B_2) \ar[r] &  \Ext^3( E_t,B_2)\ar[r] & 0 }
\end{equation}
From here it is easy to show by induction that $\Ext^{0,2,3}(E_t,B_1) =0$, 
and that 
$\dim(\Ext^1(E_t,B_1) = t$. We find that the dual quiver has $k$  meson
 arrows corresponding
to $\Ext^1(B_5,B_1) = \Ext^1(E_k,B_1)$. This is exactly the matter content
 predicted by
Kutasov \cite{kutasov}. It is very clear that we can tie
the mesons  to a particular step in the extension, so this determines
their index structure with respect to the basis $e_1, \dots e_t$.

For completeness we also need to verify that the brane $B_5$ is rigid, 
namely , that there are no massless adjoints available for brane $B_5$.
 This is 
most easily done in the field theory and will be left as an exercise for 
the reader.

Now let us analyze the superpotential of the dual theory. As we have seen 
in subsection 
\ref{sec:seidualrev} , this is best done by
examining the original theory in a quiver with branes and antibranes included 
simultaneously. For this, we will consider a theory of branes $B_1+\bar B_2 
+ B_5$, where $B_5$ is written  in terms of the original branes, 
$kB_2 + B_3$.

\begin{figure}[ht]
\begin{center}
\leavevmode
\epsfxsize=7cm
\epsfbox{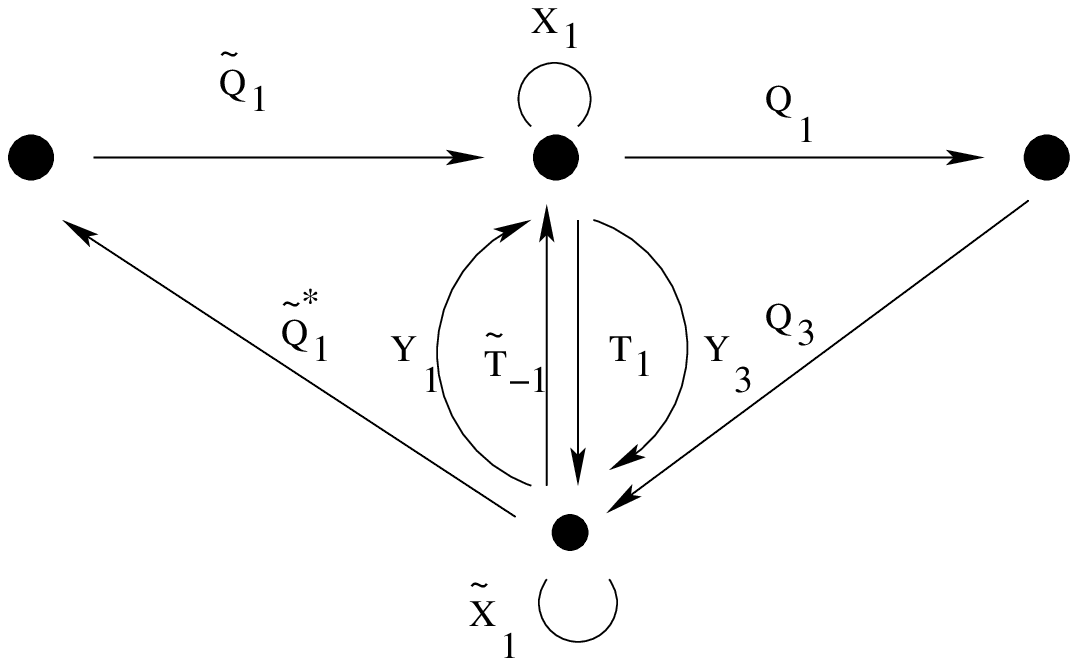}
\end{center}
\caption{Quiver with antibranes in the presence of adjoints. 
The ghost charges are explicitly labeled.}
\label{fig:quiver5.eps}
\end{figure}

The new ingredient in the construction is that we have extra arrows between
 the branes and the 
antibranes, which arise from shifting $\Ext^{1,2}(B_2,B_2)$. These are 
depicted in 
the figure \ref{fig:quiver5.eps}.

The new fields to consider are the field $Y_{1,3}$ and  $\tilde \chi$.
 One can argue via the 
graded Chern Simons theory that the term $\tr(\chi^{k+1})$ needs to be 
completed to 
a graded version of this term, which includes a term 
of the form $\tr(\tilde\chi^{k+1})$.
This is a term expected in the superpotential of the dual theory.

Similarly we find that there are terms in the potential required 
by tachyon condensation which are of the form
\begin{equation}
\tr(\chi_1 Y_1 T_1 +  T_1 Y_1 \tilde \chi_1  +  \tilde Q_1^* T_1 \tilde Q_1)
\end{equation}

In the dual theory we know that most of these fields are not part of the
 spectrum.
Some are eaten by the Higgs mechanism, and some others are eaten by masses
 generated in the 
superpotential. From this point of view, it is important to notice that
 when we give a vev to 
$\chi_1$, we generate mass terms for $T_1$ and $Y_1$ jointly. Not all of the
 terms of $T_1$ get 
a mass however, because $\chi_1$ does not have maximal rank.
Now we need to eliminate the fields that acquire a 
mass and integrate them out.

In particular here we will find that we need to satisfy the 
equations of motion for
$Y_1$, $T_1 <\chi_1>  =  \tilde \chi_1 T_1$. Once we choose  
the vacuum solution
$<\chi_1>^{(i)}_{j} \sim \delta_{i,j+1}$
\footnote{This structure is chosen by the form of the repeated 
extension that gives rise to brane 
$B_5$}, 
we find 
that $(T_1)_i = \tilde \chi_1 (T_1)_{i-1} $, so we can solve 
for all of the components of
the tachyon 
$(T_1)_{i}= (T_1)_{1} (\tilde\chi_1)^{i-1}$,
 where we are only making explicit the gauge indices
corresponding to the brane $B_2$, but not it's antibrane.
 Once we integrate out $Y,{T_1}_i$, we are left with the 
effective superpotential
\begin{equation}
\tr( \sum _i \tilde Q^i \tilde Q_1^*  \tilde\chi_1^{i-1} (T_1)_1 ) 
\end{equation}
In the above, one can follow the exact sequences to realize 
that in the dual theory we need to 
identify $\tilde (Q_1)_i \sim M_i$, $\tilde Q_1^* \sim \tilde q$, 
$(T_1)_1 \sim q$. 
Thus the above equation reproduces exactly the superpotential
 written in \cite{kutasov}, namely
\begin{equation}
W = \tr( \sum_i  q M_i \tilde q \tilde \chi^{i-1})+ 
\frac1{k+1}\tr(\tilde\chi^{k+1}) 
\end{equation}
This formula is consistent with our previous results when $k=1$:  
here, the adjoint field
has a mass term and is integrated out, this can be done before 
or after performing
the duality. 
One might worry that one needs to integrate out other fields, so that 
this procedure might spoil the form of the 
superpotential. The only dubious point
of our derivation is 
that the equation of motion of $Y_1$ that we used can contain additional 
terms. These are indeed proportional to $\chi^{s_1} Y_3 \tilde \chi^{k-s_1-1}$
and are obtained from the completion of the $\tr(\chi^{k+1})$ term so that
it is covariant with respect to the graded gauge group for the $B_2\bar B_2$
brane anti-brane system. Notice that these extra terms involve fields with 
ghost charge different than one. We expect these to have a mass of the order
of the string scale, and to be completely irrelevant in the infrared, where 
we are taking the decoupling limit $\alpha'\to 0$.

The configuration that we studied can be generalized easily to a 
situation where we have more than 
one arrow going between branes $B_1$, $B_2$ (let's say $n_{12}$ of them),
 and between brane $B_2$ and $B_3$, ($n_{23}$ of them).
The maximal extension of brane $B_3$ by branes of type $B_1$ will have 
charge 
$(0,n_{23}k,1)$. We will find that there are $n_{12}n_{23} k $ meson fields.
 The arrows that leave 
brane $\bar B_2$ in the dual quiver are reversed, and the superpotential is
 as above, summing over 
different types of quark fields.

\subsection{The general case}

Now we can consider a general case with adjoints and extra superpotential
terms, and dualize one node $E$.  For each node $B_i$ we need to
 consider in the dual theory a change of basis where $\hat B_i$ is 
obtained by the maximal canonical 
extension of node $B_i$ by nodes of type $E$. This is, we take
\begin{equation}
\Ext^1(B_i,E)^* \otimes E \to (B_i)^{(1)} \to B_i
\end{equation}
and repeat extending $(B_i)^{(k)}$ until the process stops.
If the process does not stop, then there is no dual theory. Thus the original 
theory must be such that the extension process is finite for the node we are
dualizing.

The brane charge of the bound state $\hat B_i$ is  
$[\hat B_i] = [B_i]+m_i[E]$.
If the original theory is given by $\sum N_i B_i + N_E E$, corresponding to  
a  gauge group $\prod U(N_i) \times U(N_E)$, then the dual theory will
have the gauge group
\begin{equation}
\prod U(N_i) \times U(\sum N_i m_i - N_E)
\end{equation}
in the new basis. This is just the brane charge vector written in the new 
basis.

The matter fields between the nodes in the dual theory are obtained by 
following the change of basis by chasing exact sequences in the quiver. 
This is equivalent to finding the massless spectrum of 
fermions between the bound states in the field theory between branes 
$(\hat B_i)$ and 
$(\hat B_j)$. For the spectrum between brane $\hat B_i$ and the node $\bar E$
 the arrows 
are reversed with respect to the original theory. 
This follows from chasing exact sequences. In particular 
the matter content between these two usually 
involves tachyons between the branes of type $E$ that form the 
bound state $\hat B_i$ and $\bar E$. 
Also the node $\hat E$ has the same number of adjoints and superpotential 
for adjoints than the original brane $E$. 

To construct the full superpotential of the dual theory 
one needs to consider adding the terms required by tachyon 
condensation and integrating out all massive fields.
Also one might need to covariantize the superpotential with respect to the 
graded group structure between branes and anti-branes. We have found 
no need to do that so far in the examples we have studied, so it is not clear 
that this is required beyond the introduction of the tachyon 
(super)potential. 

\medskip

To summarize, we have described a systematic derivation of Seiberg
dual theories for any quiver theory of finite representation type,
which produces an explicit superpotential and proves equivalences
of moduli spaces of supersymmetric vacua between the two theories.
Interesting examples of other dualities that can be analyzed with
these methods can be found in \cite{ASY,GK,ILS}.

\section{A more mathematical approach}

One can go on to ask to what extent all Seiberg-like dualities can be
discussed in this language.  In fact the dualities we are discussing
are known in mathematics, as ``tilting equivalences.''  They relate
algebras which are not Morita equivalent (do not have identical
representation theory) but for which large subcategories of
representations are equivalent.  We will quote Rickard's theorem,
which gives a sense in which all such dualities are tilting equivalences.

We also return to the point we mentioned in section 4, that the
supersymmetric field theory does not need an explicit CY embedding to
use these ideas.  Rather, all of the information for our discussion of
duality is already contained in a given quiver diagram and
superpotential.  The concepts we are explaining then allow direct
comparison to branes on CY, and systematic study of the question of
which quiver theories can be realized by embedding branes on CY.

\subsection{Quiver categories}

Notice that in section \ref{sec:Dbranecat} all of our manipulations
depended on formal properties of the derived category of coherent
sheaves, but made very little use of specific configurations of branes
in a given CY geometry, although we were implicitly assuming that this
was the situation we were in.  This was also true of the arguments in
section \ref{sec:chabas}, in which we abstracted the dynamics of
brane-anti-brane systems as far as possible in a less formal setting,
but we made no explcit mention of a particular Calabi-Yau geometry.

We can ask the question in a different way: where do derived
categories come from? Axiomatically, one can produce a derived
category starting from an abelian category. The simplest examples of
abelian categories are given by (subcategories of) $\Mod-A$ for some
ring $A$. Also, coherent sheaves are locally of the form $\Mod-A$
where $A$ is the local ring of holomorphic functions on some
Calabi-Yau space.  At least formally, we can build a D-brane category
if we are given a ring $R$, thus one can ask if there is a canonical
ring $A$ associated to a given supersymmetric quiver field theory.

In fact, the standard construction of an associative algebra (over $\BC$)
associated to a quiver with relations \cite{ARS,Gabriel}
provides the ring $A$ for the corresponding quiver gauge theory.
In the physics literature, this has appeared in the appendix of
\cite{DFR}, and in the work \cite{Berenstein} which considers the
problem of reconstructing singularities from the quiver data of a
quantum field theory.  We now review this construction.

\subsection{Path algebras}

We assume the usual translation, already reviewed in section 2, from
quiver diagrams to the field content and gauge symmetry of a supersymmetric 
field theory.  

The path algebra $\CP(Q)$ associated to a quiver $Q$ has the 
following generators: 
\begin{itemize}
\item A projector $P_i$ for each node $i$, satisfying $P_i^2 = P_i$. 
We also have $\sum_i P_i = 1$.
\item A generator $\phi^\alpha_{ij}$ for each arrow of the quiver from the
node $j$ to the node $i$.  (We index the arrows by $\alpha$ in this
general discussion; of course one could give names or other labels to
these generators).  This satisfies
\begin{eqnarray}
\phi^\alpha_{ij} P_k = \phi^\alpha_{ij} \delta_{jk} ; \\
P_k \phi^\alpha_{ij} = \delta_{ik} \phi^\alpha_{ij} .
\end{eqnarray}
\end{itemize}
As usual, products of these generators not constrained by the relations
we just gave are considered to be new, linearly independent elements of
the algebra.  If we have no superpotential, we are done; 
this defines the free path algebra $\CP_0(Q)$.

If we have a superpotential $W$ which is a single trace of products of
the fields, we can regard it as a cyclically symmetric function of the
noncommuting variables $\phi^\alpha$ in an obvious way.  The relevant
algebra $\CP_W(Q)$ is then defined in terms of the free path algebra
by imposing all of the F-flatness conditions $\p W/\p \phi^i=0$ as
additional relations.  Since $W$ is a single trace, these will be
linear relations, and one can phrase the result as the quotient
$$
\CP_W(Q) = \frac{\CP_0(Q)}{\CI(W'=0)}
$$
where $\CI(W'=0)$ is the two-sided ideal given as sums of
terms from $W'$ multiplied by elements of $\CP_0(Q)$ on both sides.

A field theory configuration which solves the F-flatness conditions
now provides a {\em finite dimensional} representation of this path
algebra, i.e. a map $R$ from elements of $\CP_W(Q)$ to linear
transformations acting on a vector space $V(R) \cong \BC^N$,
respecting the relations.  The operator $P_i$ defines a projection
onto a subspace $V_i(R) \cong \BC^{N_i}\in V(R)$ which is acted on by
the $U(N_i)$ gauge group, while the representation matrices of the
$\phi_{ij}^\alpha$ are the explicit matter configuration.

Normally we are only interested in complex gauge equivalence classes
of configurations, and this translates into the statement that we are
interested in (a certain class of) modules $M$ over $A$, i.e. those which
can be obtained by quotienting finite dimensional free modules.
Modules are defined more abstractly than representations, by prescribing
a set of generators and a product $A \times M\rightarrow M$, and
formulating the theory in these terms makes it far more general.
For the specific application to quiver gauge theories, however,
one can think in terms of gauge equivalence classes of representations.

\subsection{Homological algebra}

The physics reader might consider the foregoing to be obvious
restatement of what he has been doing for a long time in finding
solutions of supersymmetric gauge theory (which may bring to mind
Monsieur Jourdain's remark).  However there are some crucial
constructions one can make for the particular case of quiver
theories, which are well-known mathematically, but to learn about
them there one needs to know the translation we just gave.  

The main point is that one can easily show that representations of
a path algebra $\CP_W(Q)$ as defined above form an abelian category.
The basic definition is the following: a {\it homomorphism} between
quiver representations, $\rho \in \Hom(R,S)$,
is a linear map $\rho$ from $V_R$ to $V_S$, satisfying
\begin{equation}\label{eq:defqhom}
S(a) \rho = \rho R(a) \qquad \forall a\in\CP_W(Q).
\end{equation}
Expanding out the definitions, one finds that this is a collection
of $N_i(S) \times N_i(R)$ matrices, which live in the upper diagonal block
of a collection of $GL(N_i(S)+N_i(R))$ gauge transformations.
These should be thought of as ``partial gauge transformations'' which
preserve the configuration of the joint $U(N_i(S)+N_i(R))$ gauge theory
obtained by direct sum.

There are several axioms to check in saying that the representations
$R$ and the morphisms $\Hom(R,S)$ make up an abelian category.  All
are obvious, if we think of them as finite dimensional linear maps:
the morphisms have an associative multiplication, and they have
kernels and cokernels.  They are less obvious if one regards $\Mod-A$
as the definition and one must put appropriate conditions; for the
concrete application at hand we assume the appropriate conditions.

This is the general definition of the $\Hom(R,S)$ which we used in
section 2, for any quiver gauge theory.  What we will not explain here
(it is standard mathematics; some of this is in \cite{Trieste}) is
that this definition leads to definitions of $\Ext^p(R,S)$ and many of
the other concepts in our arguments, all the way up to the derived
category of finite dimensional representations of the quiver algebra.

Let us denote this category of representations arising from a theory
$T$ with quiver $Q_T$ and superpotential $W_T$, $\Mod-\CP_{W_T}(Q_T)$,
simply as $\Quiv T$.

In examples, these definitions of $\Ext^p(R,S)$ satisfy the simple
physical interpretations known for $p=1$ and $p=2$ from geometry,
branes and other frameworks.  In particular, $\Ext^1(R,S)$
corresponds to variations of bifundamental matter from $R$ to $S$
which is not gauge and not lifted by the superpotential, and
$\Ext^2(R,S)$ corresponds to inactive superpotential constraints.
The higher $\Ext^p(R,S)$ encode information about redundancies of
superpotential constraints (e.g. see \cite{Trieste}).

This is one underlying mathematical context for quiver gauge theory.
As we have seen, it is the natural context in which to discuss
Seiberg duality.  Let us expand a bit on this point.

In our previous discussion, we described classical dualities as the
matching of moduli spaces of supersymmetric vacua of the two theories.
Let us compare to this, the statement that this categorical structure 
matches between two quiver theories.  In other words, the set of F-flat
configurations match, and the partial gauge equivalences and higher
$\Ext$'s match.  This type of relation is known as an equivalence
between categories.

One obvious difference is that we need to talk about the D-flatness
conditions or FI terms to make the first statement, while we do not in
the second.  In fact one can find necessary and sufficient conditions
to solve the D-flatness conditions, which only depend on the
categorical structure \cite{King}.  Thus the second statement already
contains the information needed to check the first statement, for any
choice of FI terms.  In this sense, the second statement is stronger.

One can go on to ask whether equivalence of categories is stronger than
matching solutions of F-flatness in the sense of matching off-shell
information about the superpotential.  This is an interesting question
which we will not address here.

\subsection{Tilting equivalences}

In section 2, we saw that the basic mathematical concept underlying
Seiberg's original duality (for $U(N)$ and with all the caveats we
mentioned) was the reflection functor of \cite{BGP}.  This was defined
in 1973 and the mathematicians were not idle in the meantime.  

By the previous discussion, we want to rephrase the problem of
classical duality between quiver gauge theories $T$ and $T'$, as the
problem of finding an equivalence betweeen the categories of modules
of their respective path algebras.  This brings to mind Morita
equivalence: two algebras are Morita equivalent if they have
equivalent categories of modules.

However, we did not find dualities for all ranks of the gauge groups,
just a large subset.  For example, if the dual theory would have had
$N_c<0$, obviously we do not have a dual theory.  We excused this
fault in the previous discussion by saying that the original theory
would not have solved the D-flatness conditions for large FI terms,
but now we are not imposing D-flatness and we do not have this out.

In fact, the two path algebras are not Morita equivalent.  Rather, the
duality corresponds to an equivalence between two large subcategories
of $\Quiv T$ and $\Quiv T'$, not the full categories, which is weaker.

A nice introduction to the general theory of such equivalences can be
found in the textbook \cite{Konig} (which applies it to rather
different problems coming from modular representation theory of
groups).  In fact, the first example they discuss, that of
equivalences between algebras (in their notation) $A_1$ and $A_3$, is
just the original $U(N)$ Seiberg duality we discussed.

We will not go much further into the details of this, but just cite
the general results of the theory.  First of all, one wants to see
that these equivalences of large subcategories of modules, imply that
the derived categories of the two module categories are equivalent.
This was found to be true for the explicit equivalences discovered by
mathematicians and one might take it as axiomatic.  In discussing
theories which arise from D-branes, we can appeal to the topological
open string construction of \cite{DougDC} to justify this.  In any
case, the intuition is that the derived category is a universal
construction in which the formal manipulations we were doing before,
always lead to specific configurations of a dual theory, no matter
what we started with (and thus lead to a complete equivalence), while
retaining enough information to force the specific equivalences of
representations in the subcategories.

The main result in this theory is Rickard's theorem \cite{Rickard}:
any equivalence between two derived categories $D(\Mod-A)$ and
$D(\Mod-B)$, can be realized as a tilting equivalence.  There are
various definitions of tilting equivalence; let us start with the simplest,
and later give a more sophisticated one (the theorem as stated only
holds with the second definition).

A simple definition is the following: we need a tilting module
$T$, an $A$-module such that
\begin{itemize}
\item The projective dimension of $T$ is zero or one;
\item $T$ does not have self-extensions, i.e. $\Ext^i(T,T)=0$ for $i>0$; and
\item $A$ (as a module) has a finite resolution whose terms are direct summands
in some number of copies of $T$.
\end{itemize}
Clearly explaining even what this simple definition means is going to
get us into a long discussion, but physically the idea is more or less
that $T$ is a set of branes from which no bound states can form, but
if we add to them their antibranes we can generate the whole category
as bound states.

One then has a tilting equivalence between $A$ and $B=\Hom(T,T)$, the
endomorphism ring of $T$.  This is also hard to visualize physically,
but let us see how the original example works.  For this, we take
$T$ to be the direct sum
$$
T = B_1 \oplus B_3 \oplus B_6
$$
in the notations of \rfe{boundstates}.  The no bound state condition
is clear.  The endomorphism algebra is now generated by 
$\alpha\in\Hom(B_3,B_6)$ and $\beta\in\Hom(B_6,B_1)$.  These turn
out to satisfy the single relation $\beta\alpha = 0$.

This is precisely the path algebra of the Seiberg dual theory, with
the relation \rfe{Mdef}\ (but leaving out the other dual quark which we
put in to have a superpotential).  Classifying representations of this
algebra will then lead us to the brane content of this theory.

This argument strikes us as both very short and to the point, and very
mysterious.  Anyways, the tilting equivalence for this simplest example
of duality is relatively simple, so this might be a viable framework
in which to look for new dualities.

The axioms for a tilting module are rather restrictive, and to get
the general result we cited, one needs a more general definition,
involving a tilting complex.  We will not give the axioms this must
satisfy, but quote the actual transformation one gets: if $T$ is
a tilting complex, then the functor $\CF_T$ from $D(\Mod-A)$ to
$D(\Mod-B)$ defined by
$$
X \mapr^\CF X \otimes_A^L T
$$
is an equivalence (and all equivalences can be written this way).

This is rather formidable in general but does reduce to something more
concrete (at least, comparable to the first definition) in cases in
which the modules exist on both sides.  Granting the earlier point
that classical Seiberg dualities must give equivalences of derived
categories, this provides a complete answer to the problem of finding
such dualities, in principle.  However, we suspect the physics reader
who tries to follow this up will soon find himself seeking
professional help.

The main reason we cited this was to make the point that this
transformation is formally very similar to the Fourier-Mukai
transforms which describe autoequivalences of $D(\Coh M)$ for
Calabi-Yaus, again reinforcing the theme that the structures
identified in this context are more generally present in
supersymmetric quiver gauge theory.

\subsection{The Calabi-Yau condition}

It is interesting to ask what characterizes the quiver theories which
actually come out of string compactification on a Calabi-Yau.  The
most basic property of these categories of coherent sheaves (and
the derived categories) is a particularly simple form for
Serre duality (since the canonical sheaf is trivial), namely
$$
\Ext^{d-i}(S_1,S_2) \cong \Ext^i(S_2,S_1)^*
$$
on a Calabi-Yau $d$-fold.  The corresponding statement in the
derived category has been suggested as the {\it definition} of
a ``Calabi-Yau category'' \cite{Kontsevich}.

This is not a general property of quiver categories.  If there are no
relations, one can show that $\Ext^p(A,B)=0$ for all $p>1$, so it
could only be true for $d=1$.  In fact only the theory with a single
adjoint superfield would pass this test.  This is in fact the theory
describing branes on $T^2$, so imposing the Calabi-Yau condition
does lead to a sensible constraint in this case.

One might ask if it is a general property of quivers with relations
which follow from a superpotential.  One can even be more specific and
ask for $d=3$, on the grounds that ``the superpotential is a
three-form.''  This somewhat cryptic comment is justified by the idea
that relations $\frac{\p W}{\p\phi}=0$ are naturally dual to fields
(elements of $\Ext^1$), but also naturally live in $\Ext^2$.

Let us see more specifically how one might arrive at this conclusion
from a purely algebraic point of view. To do this more precisely, we will
need a working definition of $\Ext$ for modules over algebras.

The most economical definition is in terms of projective resolutions.
A projective resolution of a module $M$ is an exact sequence of modules
\begin{equation}
{\cal P}_i \mapr^{\delta_i} {\cal P}_{i-1} \to \dots \to {\cal P}_o \to M
\end{equation} 
where each ${\cal P}_i$ is a projective module for the algebra 
$\cal A$. These resolutions are also important 
to give a noncommutative definition of 
smoothness for singularities \cite{BL}.
These modules are essentially determined by a projector 
$p_i\in M_n( {\cal A})$ for some $n$. 

The defintion of $\Ext$ is then the homology of the sequence
\begin{equation}
\Ext^p(M,N) = H^p(\Hom({\cal P}, N))
\end{equation}
which can be shown to be independent of the choice of projective resolution.

The advantage of writing the $\Ext$ in this form is that we have a large class
of choices for projectors in our algebra, namely the $P_i$, so that the
modules $\CA P_i$ are projective.

Now, we also want to associate to each node in the quiver diagram a particular
representation of the group. Under the assumption that the quiver diagram 
has superpotentials of quadratic order or higher,  this is explicitly 
given 
by the following
\begin{equation}
B_{k} : ( P_i = \delta_{ik}, \phi^\alpha=0)
\end{equation}
We will choose this representation to be giving us an explicit 
module with action by ${\cal A}$ on the right. We choose this, because we
want to be able to make maps between modules by multiplication on the left
by elements of $\cal A$.

Now, we want to build a projective resolution for $B_{i}$.
Let us consider all the fields $\phi^a_{ji}$ that begin at node $i$ with $j$ 
variable. Then 
it is easy to see that we can begin to write a projective resolution in the 
following form
\begin{equation}
\oplus_{\phi^a_{ik}} (P_k {\cal A})\mapr^{\delta_1} P_i {\cal A} \to B_i
\end{equation}
where $\delta_1( a_k) = \sum ( \phi^a_{ik} a_k)$.
This map produces all the possible polynomials in the generators 
with at least one $\phi$, so 
the cokernel of this map is exactly the only term in $P_i {\cal A}$ 
that is not of this form, $P_i$.
This gives us the first two terms of the resolution, and 
one has exactness in the first term.

Now, we can try to see what relations will appear when we consider
the superpotential. To do this we consider
the following composite fields 
\begin{equation}
M^{ba}_{jik} = \phi^{b}_{ji} \phi^a_{ik}
\end{equation}
and the quantities 
\begin{equation}
W^{kij}_{ab} = \frac{\partial W}{\partial M^{ba}_{jik}}
= \frac{D_L}{D \phi^a_{ik}} \frac{\partial W}{\partial \phi^b_{ji}}
\end{equation}
where the term $\frac{D_{L,(R)}}{D \phi^b_{ji}}$ indicates that we only take a 
derivative of the equation if the field $\phi^b$ appears in the leftmost 
(rightmost) term
of the polynomial $\frac{\partial W}{\partial \phi^a_{ik}}$, which encodes 
the superpotential relations. The idea is that 
\begin{equation}
W_b' = \sum_a \phi^a W_{ab}\label{eq:resup}
\end{equation}
gives us the polynomial relations associated 
to the derivative of the superpotential with respect to the field $\phi^b$.
Thus we can try as a second term in the resolution, a sum over all 
fields that go into that node,
\begin{equation}
\oplus_{\phi^b_{ki}}( P_k{\cal A}) 
\mapr^{\delta_2} \oplus_{\phi^a_{ij}}
 (P_j {\cal A})\mapr^{\delta_1} P_i {\cal A} \to B_i
\end{equation}
The map $\delta_2$ acts by  $\delta_2(a_k) 
= ( \sum W^{kij}_{ab} a_k)$. One can easily show that 
$\delta_1\circ\delta_2 = 0$, because \rfe{resup}\ are exactly the relations 
derived from the superpotential.
Similarly we can consider the next term in the exact sequence 
\begin{equation}\label{eq:presolution}
P_i {\cal A} \mapr^{\delta_3} \oplus_{\phi^b_{ki}}( P_k{\cal A}) 
\mapr^{\delta_2} \oplus_{\phi^a_{ij}} (P_j {\cal A})
\mapr^{\delta_1} P_i {\cal A} \to B_i
\end{equation}
where $\delta_3 a = (\phi^b_{ki} a)$. Again, one can show that 
$\sum_b W_{ab}\phi^b$ are exactly the superpotential relations 
for $\phi^a$, so we obtain $\delta_2\circ\delta_3=0$.

We thus have a complex \rfe{presolution}, whose terms are projective.
For this to be a projective resolution, we need it to be exact.
This condition depends on the choice of superpotential.  For example,
it will fail for $W=0$.  It is also clear that every field must appear
in the superpotential. 

On the other hand, there are many superpotentials for which this is a
resolution.  The case which is best understood mathematically is if
the path algebra is Koszul (e.g. see \cite{Floystad}), essentially
meaning that there is a projective resolution in which the maps are
linear in the generators.  This includes the $\BC^3/\Gamma$ McKay
quiver path algebras \cite{ItoNakajima}, so in this case the CY
condition is satisfied.

A Koszul algebra must have purely quadratic relations and a purely
cubic superpotential.  On the other hand, as is well known, branes on
CY need not have cubic superpotentials, despite naive appearances
from \rfe{holocs}.  Thus we do not assume this in formulating the CY
condition.  This point is explained in
\cite{Lazaroiu4,Tomasiello,DGJT} and will be explained further
elsewhere.

Given that \rfe{presolution}\ is a projective resolution,
we believe the quiver category has Serre duality, and that this
is a CY category
(we are not at this point claiming that this is necessary).
It is easy to check
that the morphisms between fractional branes (simple representations) 
correspond to those we used in section 4, and satisfy Serre duality.

A bit more explicitly, one first checks that
$\dim(\Hom(P_i{\cal A}, B_k)) = \delta_{ik}$. If the superpotential is 
cubic or higher order then each term 
in the $\delta_{\alpha}$ will be mulitplying 
by polynomials in the $\phi$ of degree equal to one or higher, so in
 the homology 
sequence of
of $\Hom([{\cal P} \to B_k],B_j)$ all the chain maps are 
zero.
With these conditions one finds that 
$dim(\Ext^0(B_i,B_k) = \delta_{ik}$, $dim(\Ext^1(B_i,B_k) = n_{ki}$,
$dim(\Ext^2(B_i,B_k))= n_{ik}$, $dim(\Ext^3(B_i,B_k) = \delta_{ik}$;
where $n_{ki}$  are the number of arrows in the quiver begining
at node $i$ and ending at node $k$.

One can now ask when do we expect the sequence above to have
homology, so that it is not a resolution.  The simplest example is
$W=0$, for which $\delta_2=0$ and there is homology in the third term
of the resolution.

A simple argument also shows that if the 
complex written  above is a resolution for every $B_k$
then 
the algebra needs to be infinite dimensional
over the complex numbers. To show this we need to 
consider the brane given by $B= \oplus B_k$ over all
posssible elementary fractional branes. If the
algebra is finite dimensional, then all of the terms of the resolutions
have finite 
dimension. This permits us to 
show that the (virtual) dimension of the resolution of $B$ , and
hence the dimension of $B$ is 
zero. This follows because $P_i{\cal A}$  appears repeated as 
each term is counted for $Ext^{1,2}$
either when arrows depart from a node, or when they arrive at the node. For
$Ext^{0,3}$ this pairing is obvious.

For quotient singularities the algebras involved are always
infinite dimensional, as their center is the ring of the 
quotient variety \cite{BL}, and it is indeed the case that 
in this situation the algebra satisfies the CY condition.

\subsection{Non CY examples}

Not all of the theories we considered in section 4 satisfy the CY
condition.  For example, the original SQCD had no superpotential.  On
the other hand, it can be obtained by wrapping branes on CY.  

It is also true that ``tilting equivalence'' and many of the other
structures make sense for general quivers, not just CY quivers.  On
the other hand one does not expect that in general the dual algebra
can be described using a superpotential; the relations might be more
complicated.  

One might conjecture that when this can be done, it is
because the original quiver is a subquiver of a larger quiver which
does satisfy the CY property, meaning that by eliminating some of the
nodes and all arrows to these nodes one reduces the larger quiver to
the smaller quiver.  One is not allowed to eliminate arrows between
nodes which remain in the subquiver, so this relation is {\it a priori}
nontrivial.

\begin{figure}[ht]
\begin{center}
\leavevmode
\epsfxsize=7cm
\epsfbox{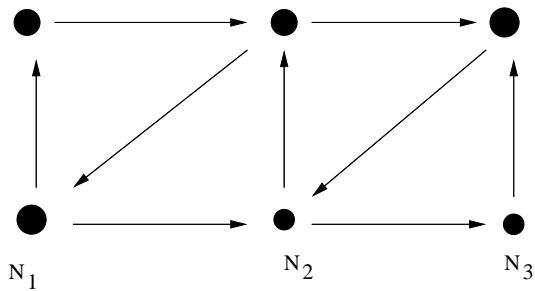}
\end{center}
\caption{$\BZ_n\times \BZ_m$ quiver diagram: the pattern repeats
itself to get an $n\times m$ periodic lattice}
\label{fig:cnm.eps}
\end{figure}

For example, one can embed the SQCD quiver in the $\BC^3/{\BZ_m\times
\BZ_n}$ McKay quiver, as in figure 6.  One sets all $N_i=0$ except for
$N_1, N_2, N_3$ as marked on the figure.

We must now distinguish the $\Ext$ groups for the small algebra and
the larger CY algebra; these are in general different because extra
nodes appear as data in the projective resolution of the fractional
branes in the big quiver.  Our arguments strictly speaking (and
whenever we discuss superpotentials) apply to the larger quiver.

This raises another general question: can all quivers can be embedded
in CY quivers; if not, what is the constraint under which this is
possible.

\subsection{Partial resolutions}

Many singularities can be built from partial resolutions
of an orbifold singularity, giving us access to a large class of 
field theories other than orbifolds. These techniques have been used 
to geometrically engineer other singularities and make them tractable 
from a D-brane analysis; and have led to the notion of toric duality 
\cite{FHH, FHH2, FHHU, FFHH, BP} by arriving to different
field theories associated to the same singularity 
by taking different paths to get there. Here we will explain how 
this procedure of partial resolutions can be used in a more abstract 
setting.

The procedure for generating these partial resolutions is as follows
(see for example \cite{FHH, BP})
\begin{itemize}
\item Give large FI terms to some nodes in a known quiver diagram.
\item One decides that some fields (arrows) in the quiver
 get large vevs. 
\item One reduces to the unbroken gauge group. This fuses the nodes that 
are connected by large vevs into one.
\item Integrate out massive fields.
\end{itemize}

We have seen very similar procedures in our discussion of duality.
We have turned on large FI terms to be able to go around singularities,
turning on large vevs forces branes to form bound states, and the 
procedure of integrating out massive fields (which can include tachyons)
gives rise to the superpotentials of the dual theory. Thus all of these 
operations fit naturally into the discussion of this paper.

We want to argue that the procedure above is like localization and that 
it can be described in more algebraic terms. The most important point in 
the above is the second one. We choose some fields to get large vevs and 
we force certain branes to form bound states, so we should throw away 
certain brane configurations where this does not happen. 

In a geometrically engineered theory,  the 
branes we will throw away have wrapping number
around the cycle that is resolved, so they are not local branes at 
the leftover singularity. At an abstract level however, we need to be able 
to make these statements without geometry.

Giving large vevs to some fields forces the arrows we condense
$\phi^a_{ij}$ to give isomorphisms between certain gauge groups
and they break the gauge group to a common diagonal.

At the level of representation theory of the original quiver
this means we need to focus on
brane configuratons where $N_i = N_j$. The vev for $\phi^a_{ij}$ being 
large means that it is essentially invertible. This is, one can find in each 
of these representations a matrix $\beta^a_{ji}$ such that
\begin{eqnarray}
\beta^a \phi^a &=& P_j\\
\phi^a \beta^a &=& P_i
\end{eqnarray}
and we can always choose $\beta$ such that  
$\beta^a P_k = \delta_{ik} \beta^a$ and $P_k \beta^a = 
\delta{jk} \beta^a$. If the original $\beta$ does not satisfy this 
condition then $\beta' = \beta\phi\beta$ does.

Let $\cal A$ be the original quiver algebra. The representations above are 
representations of the algebra where we adjoin $\beta$ as a generator 
of the algebra with the relations imposed above,
 let us call this bigger algebra ${\cal A}[\beta]$.

Since we have an embedding of algebras 
${\cal A} \subset {\cal A}[\beta]$, every representation of the latter 
algebra is automatically a representation of ${\cal A}$.
Given $R_1, R_2$ representations of ${\cal A}[\beta]$ one can prove 
by a matrix argument that 
$Hom_{\cal A} (R_1, R_2) = Hom_{{\cal A}[\beta]}(R_1,R_2)$.
In particular kernels and cokernels of maps between representations
of $\cal A[\beta]$ will be also such type of representations.
Thus, if we consider the category of the representations of
$\cal A[\beta]$ as a subcategory of the representatoins of $\cal A$, 
then this category is closed under taking kernels and cokernels.

Also one can prove that given any short 
exact sequence $0\to R_1\to R\to R_2\to 0$
of $\cal A$ modules, where $R_1$ and $R_2$ are 
${\cal A}[\beta]$ modules, then $R$ is a representation of 
${\cal A}[\beta]$ as well. 

The category $Mod({\cal A}[\beta])$ is therefore a full 
(abelian) subcategory of $Mod({\cal A})$: it is closed
under the operation of taking kernels, cokernels. The $Hom$
coincide, and the category is closed under extensions. 
The above argments have shown 
that 
$Ext^{0,1}_{\cal A} (R_1, R_2) = 
Ext^{0,1}_{{\cal A}[\beta]}(R_1, R_2)$, so that the matter content between
bound states of branes in the partial resolution is completely equivalent to
the matter content between the same bound states as seen from the original 
quiver diagram. 
The process of integrating out massive fields is exactly what
is captured by the homological algebra of the $Ext$ functors.

In the end we should have 
$Ext^i_{\cal A}(R_1, R_2)= Ext^i_{{\cal A}[\beta]}(R_1, R_2)$.
This is exactly the behavior that one expects for the $Ext$ functors
in an algebraic geometry for 
coherent sheaves (branes)  whose support is entirely contained 
in an open set (localization)  of some known geometry. 

In particular, if some quiver algebra satisfies the CY condition, then any 
partial resolution of the algebra as given above should
also satisfy the CY condition.

This also poses the question of how the duality 
transformations between a quiver and it's partial 
resolutions are related.

Geometrically, the partial resolution produces a space $X$, which can
be further resolved to produce $\hat X$ which embeds into $Y$,
where $Y$ is a resolution of $\BC^3/\Gamma$. 
The process of going from 
$\BC^3/\Gamma$ to $X$ blows up some cycles to 
infinite size and thus reduces the 
local K-theory group (the coherent sheaves of compact support).
The quiver category $D(\Mod-Q_\Gamma) \cong D(\Mod Y)$ via the Mckay
correspondence. The classes (of compact support)
that dissapear 
geometrically are the classes that are
wrapped on the blown up cycles. These are identified in
the field theory as the 
classes for which $N_i\neq N_j$.

We have the inclusion $D(Mod Q_T)\subset D(\Mod-Q_\Gamma)$, as we
described above, for the quiver category of the partial resolution
field theory, and $D(\Mod \hat X) \subset D(\Mod Y)$. These are both
subcategories of the same category, and they span the same K-theory
lattice, so these two coincide. We can then identify $D(\Mod Q_T)
\cong D(\Coh \hat X)$. This will hold whenever one can resolve
completely the singularity. This is the case for toric singularities:
turning on generic D-terms resolves all the singularities, and $X$ is
the moduli space of the field theory with $N_i=1$ for all the nodes of
$Q_T$.  It should be said that there are situations in which this
condition will not hold, e.g. on orbifolds with discrete torsion.

\subsection{Toric duality is generalized Seiberg duality}

It is not true that all partial resolutions of a quiver theory are
Seiberg dual.  Rather, the claim of \toricrefs\ is that two partial
resolutions of the same orbifold theory which lead to theories $T_1$
and $T_2$ which describe the same space $X$, i.e. the moduli space of
supersymmetric configurations with all $N_i=1$ is $X$ in both cases,
are Seiberg dual.

The previous considerations give us a very short argument to this
effect.  As we discussed, partial resolution of the orbifold theory
$\Gamma$
leads to a gauge theory $T$ such that $D(\Mod-Q_T) \cong D(\Coh X)$.
The assumption that two such theories produce the same $X$ then
implies that $D(\Mod-Q_{T_1}) \cong D(\Mod-Q_{T_2})$.  

By Rickard's theorem, any Seiberg duality would have to induce such an
equivalence.  In this sense, toric duality must be a generalized
Seiberg duality -- it will be a tilting equivalence, but it is not
obvious whether the equivalence can be realized by some sequence of
Seiberg dualities, each of which acts on a single node.
In the examples of \toricrefs, this tended to be true, but there
might be more general dualities which could be obtained this way.

\section{Conclusions and further directions}

Seiberg duality is an important aspect of $\CN=1$ supersymmetric gauge
theory.  After its original discovery in supersymmetric QCD, it has
been generalized to a very large class of theories with multiple gauge
groups and varied matter content.

There have been various derivations, each of which works in some class
of theory and suggests some underlying origin of the phenomenon.  The
arguments of \cite{APS}  apply to deformed $\CN=2$
theories.  Suspended brane arguments \cite{HW,GK,EGK}
work in theories that can
be realized as brane webs, a certain class of quiver theories and
orientifolded quiver theories.  These seem to be a special case of the
larger class of quiver theories obtained by placing branes near a
partially resolved orbifold singularity.  This allows more contact
with geometry but is still local on the CY.  Not all combinations of
cycles in a CY can be so realized, and one can further extend the
class of geometrically engineered theories by allowing more general
combinations of cycles.  In the stringy regime, the very definition of
cycle has to be generalized \cite{DougDC,AD}.  Eventually this line of
development merges with the general discussion of $\CN=1$ string/M
theory compactification.  On the other hand, it seems unlikely that
all $\CN=1$ theories arise from string and M theory, making one wonder
whether such explanations are somehow missing some simpler point.

All of these constructions work by embedding into quantum theories
with some independent definition and thus provide some explanation of
the duality at the quantum level.  This does not necessarily mean that
one can get the explicit quantum effective theory, however.  The best
understood cases remain those in which quantum corrections are either
absent or operate by destabilizing certain supersymmetric vacua, while
other components of the moduli space are the same as in the classical
limit.

In this work, we gave a rather different type of argument: namely, we
abstracted the essential features of the brane arguments at the
classical level, and discussed how they could be applied to any quiver
theory.  The essential idea is that in a given ``geometry,'' one can
consider the set of all possible supersymmetric configurations of
branes and antibranes.  This set can be described as the moduli space
of vacua of the combined world-volume theory of the branes, but making
such an explicit description requires a choice of {\it basis}, a
finite set of elementary branes in terms of which each configuration
can be constructed in one way.  There are many possible choices of
basis, and each leads to a different gauge theory with the same set of
configurations.  The relation between two such descriptions is Seiberg
duality.

Making this precise requires being able to work with branes and
antibranes at the same time, which is not possible in conventional
supersymmetric gauge theory.  This is why previous arguments relied on
particular realizations of the branes in string theory.  However, in
the context of quiver theories, one can make a more abstract
definition of antibrane in terms of its defining property, that
tachyon condensation between a brane and its antibrane leads to the
vacuum configuration, using homological algebra techniques.  This
agrees with the string theory definition coming from D-branes when
this makes sense, but does not presuppose that the theory can actually
be realized in terms of D-branes.  Indeed, these arguments do not make
use of any conventional geometric definition of a space with branes in
it, suggesting that quantum Seiberg duality does not require such a
picture either.\footnote{
One can regard the quiver point of view as a working definition of
``noncommutative geometry.''}

It is a very interesting question, which quiver theories have a
geometric interpretation, meaning that they can be realized by
D-branes on a Calabi-You threefold, and which do not.  It is hard
to believe that all of them do, and if this were true, it would
probably have dire consequences for the predictive power of string
theory.  Assuming that it is not true, one would like to know the
necessary and sufficient criteria for a geometric realization.

One test which might be applied is that the center of the algebra
should be a Calabi-Yau threefold (in some sense).  Another test is
that the category of quiver representations must have homological
dimension $3$ and admit Serre duality.  Both conditions come directly
out of the theory of branes on CY$_3$ but are not at this point
obvious for general quiver theories.  In fact, we believe that the
second condition is rather weak and will return to this point in
subsequent work.

Of course we know of phenomena such as discrete torsion that require
generalizing our idea of ``geometry'' in any case, and there might be
non-geometric string theory compactifications as well.  However, it is
hard to formulate an accessible question about which theories can come
from branes in string theory at this point.

Comparing to the work of \cite{FHH, FHH2, FHHU, FFHH, BP} on toric
duality, our arguments seem to us conceptually simpler than basing the
discussion on an underlying geometry.  In particular, they completely
bypass the complicated question of deriving the quiver from the
geometry.  On the other hand, we believe one could obtain a fairly
complete explanation of toric duality from this point of view, based
on the framework of the generalized McKay correspondence, which
provides a direct (though somewhat abstract) relation between the
geometry of orbifold resolution and quivers.  The outline of this is
already present in existing mathematical work -- the various partial
resolutions of the orbifold lead to birationally equivalent spaces,
which are connected by performing flops; each such flop can be
realized as a known transformation on the derived category which would
be a tilting equivalence between the categories of quiver
representations on both sides.  It will be interesting to develop this
understanding.

Perhaps the most interesting direct application of the relation to
tilting equivalences is that it provides a way to look for new Seiberg
dualities, not realized as a succession of the known dualities which
act on a single node.  Much is known about the case of ``tame''
algebras (i.e. finitely many indecomposable representations), which
might be relevant.

It is interesting to note that the key step in the derivation, namely
\rfe{gaugetrans}, is formally a gauge transformation on the larger
system of branes and antibranes we are using to represent a
configuration.  In this sense, we find that Seiberg duality is itself
a gauge symmetry.  This observation can be pushed further; the key
step leading to the derived category framework in which these
considerations naturally fit, is to allow equivalences involving
brane-antibrane annihilation (quasiisomorphisms) to play the same role
as the original gauge transformations (homomorphisms).  In this sense,
brane-antibrane annihilation becomes a sort of generalized gauge
equivalence.  Although we are seeing it here in the specific context
of quiver theories, it seems to us that this idea should be valid more
generally, and could potentially be very fruitful.

We only derived equivalences between classical moduli spaces, but we
believe this is the most nontrivial test of the duality and that this
captures the heart of the phenomenon.  We made some comments in
section 2.6 about issues related to quantum effects in four
dimensions.  Another consequence/test of the idea that the origin of
the duality is effectively classical, is that similar dualities should
exist for quantum theories in dimensions 1, 2 and 3 (and even 0 if one
considers matrix integrals).  This is known in 3 dimensions and indeed
many brane constructions do extend to this case \cite{GK}.  From the
point of view here, it would be particularly interesting to study
dimension 1 (quantum mechanics) as in this case one is not comparing
moduli spaces.  It seems very plausible that these dualities would act
as symmetries on the BPS spectrum; an example (with eight
supercharges) in which this is known to be true is the action of the
Weyl group on the affine Lie algebras of \cite{HM}.

It is important to try to push this understanding beyond quiver
theories.  A next step in complexity is to incorporate the other
classical groups.  As is well known, these can be expressed in quiver
language by starting with a $U(N)$ theory and restricting to fixed
points of a $\BZ_2$ action on the fields (the usual orientifold
construction).  One can in principle represent exceptional groups and
more general matter by imposing analogous non-linear conditions; the
value of this is not yet clear.  We would not claim at this point that
all phenomena in string/M theory can be understood in terms of branes,
but we do believe that branes have much more to teach us.

\section*{Acknowledgements}

D. B. would like to thank 
J. Sonnenschein, S. Katz, N. Seiberg, C. Vafa, and E. Witten
for various discussions. 
M.R.D. would like to thank O. Aharony, A. Hanany and A. King 
for discussions.

The work of D.B. was supported in part by 
DOE grant DE-FG02-90ER40542. The work of M.R.D. was supported in part
by DOE grant DE-FG02-96ER40959.


\begin{thebibliography}{99}

\bibitem{ASY}
O.~Aharony, J.~Sonnenschein and S.~Yankielowicz,
``Flows and duality symmetries in N=1 supersymmetric gauge theories,''
Nucl.\ Phys.\ B {\bf 449}, 509 (1995)
[arXiv:hep-th/9504113].

\bibitem{APS}
P.~C.~Argyres, M.~Ronen Plesser and N.~Seiberg,
``The Moduli Space of N=2 SUSY {QCD} and Duality in N=1 SUSY {QCD},''
Nucl.\ Phys.\ B {\bf 471}, 159 (1996)
[arXiv:hep-th/9603042].

\bibitem{AD}
P.~S.~Aspinwall and M.~R.~Douglas,
``D-brane stability and monodromy,''
JHEP {\bf 0205}, 031 (2002)
[arXiv:hep-th/0110071].

\bibitem{AL}
P.~S.~Aspinwall and A.~E.~Lawrence,
``Derived categories and zero-brane stability,''
JHEP {\bf 0108}, 004 (2001)
[arXiv:hep-th/0104147].

\bibitem{ARS}
{\it Representation Theory of Artin Algebras},
M. Auslander, I. Reiten and S. O. Smal\/o,
Cambridge Univ. Press, 1995.

\bibitem{BP}
C.~E.~Beasley and M.~R.~Plesser,
``Toric duality is Seiberg duality,''
JHEP {\bf 0112}, 001 (2001)
[arXiv:hep-th/0109053].

\bibitem{Beilinson}
A. A. Beilinson, 
``Coherent sheaves on $P^n$ and   problems of linear algebra'',
  {\it Funct. Anal. Appl.} {\bf 12} (1978) 214--216.

\bibitem{Berenstein}
D.~Berenstein,
``Reverse geometric engineering of singularities,''
JHEP {\bf 0204}, 052 (2002)
[arXiv:hep-th/0201093].

\bibitem{BL}
D.~Berenstein and R.~G.~Leigh,
``Resolution of stringy singularities by non-commutative algebras,''
JHEP {\bf 0106}, 030 (2001)
[arXiv:hep-th/0105229].

\bibitem{BGP}
I.N. Bernstein, I.M. Gelfand and V.A. Ponomarev,
{\it Coxeter functors and Gabriel's theorem,}
Russian Math. Surveys 28 (1973), 17--32,
(and reprinted by the LMS).

\bibitem{BS}
J.~H.~Brodie and M.~J.~Strassler,
``Patterns of duality in N = 1 SUSY gauge theories or: 
Seating  preferences of theater-going non-Abelian dualities,''
Nucl.\ Phys.\ B {\bf 524}, 224 (1998)
[arXiv:hep-th/9611197].

\bibitem{BDLR}
I.~Brunner, M.~R.~Douglas, A.~E.~Lawrence and C.~Romelsberger,
``D-branes on the quintic,''
JHEP {\bf 0008}, 015 (2000)
[arXiv:hep-th/9906200].

\bibitem{CFIKV}
F.~Cachazo, B.~Fiol, K.~A.~Intriligator, S.~Katz and C.~Vafa,
``A geometric unification of dualities,''
Nucl.\ Phys.\ B {\bf 628}, 3 (2002)
[arXiv:hep-th/0110028].

\bibitem{CKV}
F.~Cachazo, S.~Katz and C.~Vafa,
``Geometric transitions and N = 1 quiver theories,''
arXiv:hep-th/0108120.
\bibitem{CIV}
F.~Cachazo, K.~A.~Intriligator and C.~Vafa,
``A large N duality via a geometric transition,''
Nucl.\ Phys.\ B {\bf 603}, 3 (2001)
[arXiv:hep-th/0103067].


\bibitem{Diacmono}
D.~E.~Diaconescu,
``D-branes, monopoles and Nahm equations,''
Nucl.\ Phys.\ B {\bf 503}, 220 (1997)
[arXiv:hep-th/9608163].

\bibitem{Diaconescu}
D.~E.~Diaconescu,
``Enhanced D-brane categories from string field theory,''
JHEP {\bf 0106}, 016 (2001)
[arXiv:hep-th/0104200].

\bibitem{DotMan}
G. Dotti and A. Manohar,
``Anomaly matching conditions and the moduli space of
supersymmetric gauge theories,''
Nucl. Phys. {\bf B518} 575 (1998), hep-th/9710024.

\bibitem{DougDC}
M.~R.~Douglas,
``D-branes, categories and N = 1 supersymmetry,''
arXiv:hep-th/0011017.

\bibitem{DFR}
M.~R.~Douglas, B.~Fiol and C.~R\"omelsberger,
``The spectrum of BPS branes on a noncompact Calabi--Yau'', 
[arXiv:hep-th/0003263].

\bibitem{DGM}
M.~R.~Douglas, B.~R.~Greene and D.~R.~Morrison,
``Orbifold resolution by D-branes,''
Nucl.\ Phys.\ B {\bf 506}, 84 (1997)
[arXiv:hep-th/9704151].



\bibitem{DGJT}
M.~R.~Douglas, S.~Govindarajan, T.~Jayaraman and A.~Tomasiello,
``D-branes on Calabi-Yau manifolds and superpotentials,''
[arXiv:hep-th/0203173].


\bibitem{DM}
M.~R.~Douglas and G.~W.~Moore,
``D-branes, Quivers, and ALE Instantons,''
[arXiv:hep-th/9603167].

\bibitem{Trieste}
M.~R.~Douglas, to appear in the proceedings of the 2000 Trieste spring school,
and on the web.

\bibitem{EGKRS}
S.~Elitzur, A.~Giveon, D.~Kutasov, E.~Rabinovici and A.~Schwimmer,
``Brane dynamics and N = 1 supersymmetric gauge theory,''
Nucl.\ Phys.\ B {\bf 505}, 202 (1997)
[arXiv:hep-th/9704104].
\bibitem{EGK}
S.~Elitzur, A.~Giveon and D.~Kutasov,
``Branes and N = 1 duality in string theory,''
Phys.\ Lett.\ B {\bf 400}, 269 (1997)
[arXiv:hep-th/9702014].

\bibitem{FHH}
B.~Feng, A.~Hanany and Y.~H.~He,
``D-brane gauge theories from toric singularities and toric duality,''
Nucl.\ Phys.\ B {\bf 595}, 165 (2001)
[arXiv:hep-th/0003085].

\bibitem{FHH2}
B.~Feng, A.~Hanany and Y.~H.~He,
``Phase structure of D-brane gauge theories and toric duality,''
JHEP {\bf 0108}, 040 (2001)
[arXiv:hep-th/0104259].



\bibitem{FHHU}
B.~Feng, A.~Hanany, Y.~H.~He and A.~M.~Uranga,
``Toric duality as Seiberg duality and brane diamonds,''
JHEP {\bf 0112}, 035 (2001)
[arXiv:hep-th/0109063].


\bibitem{FFHH}
B.~Feng, S.~Franco, A.~Hanany and Y.~H.~He,
``Symmetries of toric duality,''
[arXiv:hep-th/0205144].

\bibitem{Fiol}
B.~Fiol, ``Duality cascades and duality walls,''
[arXiv:hep-th/0205155].

\bibitem{Floystad}
G.~Floystad,
``Koszul duality and equivalences of categories,''
[arXiv:math.RA/0012264].

\bibitem{Gabriel}
{\it Representations of Finite Dimensional Algebras},
P. Gabriel and A. Roiter, Springer-Verlag 1992.

\bibitem{GM}
S.~I. Gelfand and Y.~I. Manin,
 {\it Homological Algebra}, Encyclop\ae dia of Mathematical
  Sciences~{38},
 Springer, 1994.

\bibitem{GK}
A.~Giveon and D.~Kutasov,
``Brane dynamics and gauge theory,''
Rev.\ Mod.\ Phys.\  {\bf 71}, 983 (1999)
[arXiv:hep-th/9802067].

\bibitem{GH}
P. Griffiths and J. Harris,
 {\it Principles of Algebraic Geometry},
 John Wiley \& Sons, Inc. 1994.

\bibitem{HW}
A.~Hanany and E.~Witten,
``Type IIB superstrings, BPS monopoles, and three-dimensional gauge  dynamics,''
Nucl.\ Phys.\ B {\bf 492}, 152 (1997)
[arXiv:hep-th/9611230].

\bibitem{HM}
J. Harvey and G. Moore, ``On the algebras of BPS states,''
{\it Comm. Math. Phys.} 197 (1998) 489-519, hep-th/9609017.

\bibitem{He}
Y.-H. He, 
``Some remarks on the finitude of quiver theories,''
hep-th/9911114.

\bibitem{ILS}
K.~A.~Intriligator, R.~G.~Leigh and M.~J.~Strassler,
``New examples of duality 
in chiral and nonchiral supersymmetric gauge theories,''
Nucl.\ Phys.\ B {\bf 456}, 567 (1995)
[arXiv:hep-th/9506148].

\bibitem{ItoNakajima}
Y. Ito and H. Nakajima,
``McKay correspondence and Hilbert schemes in dimension three,''
[arXiv:math.AG/9803120].

\bibitem{King}
A.~D.~King, ``Moduli of Representations of Finite Dimensional
Algebras'', Quart. J. Math. Oxford (2), 45 (1994), 515-530.

\bibitem{Konig}
S. K\"onig and A. Zimmermann,
{\it Derived Equivalences for Group Rings,}
Lecture Notes in Mathematics volume 1685,
Springer, 1998.

\bibitem{Kontsevich}
M.~Kontsevich, 1998 lectures at the Ecole Normale Superieure.

\bibitem{kutasov}
D.~Kutasov,
``A Comment on duality in N=1 supersymmetric nonAbelian gauge theories,''
Phys.\ Lett.\ B {\bf 351}, 230 (1995)
[arXiv:hep-th/9503086].
\cite{KS}
\bibitem{KS}
D.~Kutasov and A.~Schwimmer,
``On duality in supersymmetric Yang-Mills theory,''
Phys.\ Lett.\ B {\bf 354}, 315 (1995)
[arXiv:hep-th/9505004].

\bibitem{KSS}
D.~Kutasov, A.~Schwimmer and N.~Seiberg,
``Chiral Rings, Singularity Theory and Electric-Magnetic Duality,''
Nucl.\ Phys.\ B {\bf 459}, 455 (1996)
[arXiv:hep-th/9510222].

\bibitem{Lazaroiu1}
C.~I.~Lazaroiu,
``Generalized complexes and string field theory,''
JHEP {\bf 0106}, 052 (2001)
[arXiv:hep-th/0102122].

\bibitem{Lazaroiu4}
C.~I.~Lazaroiu,
``String field theory and brane superpotentials,''
JHEP {\bf 0110}, 018 (2001)
[arXiv:hep-th/0107162].

\bibitem{morpless}
D.~R.~Morrison and M.~R.~Plesser,
``Non-spherical horizons. I,''
Adv.\ Theor.\ Math.\ Phys.\  {\bf 3}, 1 (1999)
[arXiv:hep-th/9810201].

\bibitem{Osborne}
M.~S.~Osborne,
{\it Basic homological algebra},
Graduate Texts in Mathematics,
Springer (2000)

\bibitem{Rickard}
J. Rickard,
``Morita theory for derived categories,''
J. London Math. Soc. 39 (1989), 436--456.

\bibitem{Seiberg}
N.~Seiberg,
``Electric - magnetic duality in supersymmetric nonAbelian gauge theories,''
Nucl.\ Phys.\ B {\bf 435}, 129 (1995)
[arXiv:hep-th/9411149].


\bibitem{Sen}
A.~Sen,
``Tachyon condensation on the brane antibrane system,''
JHEP {\bf 9808}, 012 (1998)
[arXiv:hep-th/9805170].

\bibitem{Tomasiello}
A.~Tomasiello,
``A-infinity structure and superpotentials,''
JHEP 0109 (2001) 030; 
[arXiv:hep-th/0107195].

\bibitem{Vafa}
C.~Vafa,
``Brane/anti-brane systems and $U(N|M)$ supergroup,''
[arXiv:hep-th/0101218].

\bibitem{Wittentop}
E.~Witten,
``Chern-Simons Gauge Theory as a String Theory,'' 
[arXiv:hep-th/9207094].



\end{thebibliography}
\end{document}